\documentclass{jfm}

\usepackage{graphicx}
\usepackage{bm}
\usepackage{amsmath}
\usepackage[utf8]{inputenc}
\usepackage[T1]{fontenc}
\usepackage{nomencl}
\usepackage{graphicx}% Include figure files
\usepackage{dcolumn}% Align table columns on decimal point
\usepackage{bm}% bold math
\usepackage[export]{adjustbox}
\usepackage[pagewise]{lineno}
\usepackage{xcolor}
\usepackage{subcaption}
\usepackage[export]{adjustbox}
\usepackage{float}
\usepackage{cancel}
\usepackage{tabularx}
\usepackage{hyperref}

\newcommand{\BibitemShut}[1]{}
\title{Optimized Flow Control based on Automatic Differentiation in Compressible Turbulent Channel Flows}

\author{Wenkang Wang\aff{1}and Xu Chu\aff{2}
  \corresp{\email{x.chu@exeter.ac.uk}}}

\affiliation{
\aff{1} International Research Institute for Multidisciplinary Science, Beihang University, 100191 Beijing, China

\aff{2} Faculty of Environment, Science and Economy, University of Exeter, Exeter EX4 4QF, United Kingdom
}

\begin{document}
\maketitle
\begin{abstract}
This study presents an automatic differentiation (AD)-based optimization framework for flow control in compressible turbulent channel flows. We developed a fully differentiable boundary condition framework that allows for the precise calculation of gradients with respect to boundary control variables. This facilitates the efficient optimization of flow control methods. The framework's adaptability and effectiveness are demonstrated using two boundary conditions: opposition control and tunable permeable walls. Various optimization targets are evaluated, including wall friction and turbulent kinetic energy (TKE), across different time horizons. In each optimization, there were around $4\times10^4$ control variables and $3\times10^{9}$ state variables in a single episode.
Results indicate that TKE-targeted opposition control achieves a more stable and significant reduction in drag, with effective suppression of turbulence throughout the channel. In contrast, strategies that focus directly on minimizing wall friction were found to be less effective, exhibiting instability and increased turbulence in the outer region. The tunable permeable walls also show potential to achieve stable drag reduction through a `flux-inducing' mechanism. This study demonstrates the advantages of AD-based optimization in complex flow control scenarios and provides physical insight into the choice of quantity of interest for improved optimization performance.
\end{abstract}

%----------------------------------------------------------------------------%
\section{Introduction}

Flow control and optimization has long been a fundamental area of research in fluid mechanics \cite[]{fukagata2024turbulent,brunton2015closed,vinuesa2024perspectives}, driven by the need to reduce drag and regulate heat transfer in engineering applications. From aerospace to automotive systems, as well as energy and process industries, controlling the flow in a manner that optimizes performance can significantly improve efficiency and reduce operational costs. 
Traditional methods of flow control are typically categorized into passive and active approaches. Passive methods, such as riblets \cite[]{garcia2011drag}, surface roughness \cite[]{yang2023prediction}, or porous media \cite[]{wang2021information,wang2022spatial,wang2021anassess,Rosti.2015}, alter the flow in a fixed manner without requiring external energy. While passive strategies are effective in certain scenarios, they are limited by their inability to adapt to changing flow conditions.
Active flow control, by contrast, provides dynamic manipulation of the flow through external actuators, allowing for more flexibility in achieving desired flow behaviors. 
An example of active control often referenced is opposition control \cite[]{Choi_Moin_Kim_1994,bewley2001dns,kametani_fukagata_2011}, which involves implementing local wall blowing and suction to negate the fluctuations in wall-normal velocity at a specific height from the wall. 
\cite{wang2024opposition} performed high-resolution large-eddy simulations (LES) to evaluate opposition control for turbulent boundary layers on wing surfaces, analyzing drag reduction and turbulence dynamics under adverse pressure gradients. Their findings highlight that its effectiveness in reducing friction drag is challenged by increased wall-normal convection from stronger gradients, especially in complex geometries such as those found in wing applications.

%\cite{sonoda2023reinforcement} applied reinforcement learning to develop advanced control strategies for reducing skin friction drag in fully developed turbulent channel flow, achieving drag reduction rates up to 37\%. This study highlights reinforcement learning’s potential over traditional methods, enabling adaptive and nonlinear control strategies that surpass conventional opposition control’s efficacy.
In the compressible regime, flow control and optimization become even more critical, as the aerodynamic and thermal challenges intensify with increasing Mach numbers. \citet{kametani2017drag} investigated the drag reduction capabilities of uniform blowing in supersonic wall-bounded turbulent flows, concluding that Mach number dependence primarily stems from varying thermal properties such as density and temperature, similar to the effect of Mach number on turbulent statistics in uncontrolled flows. \citet{yao2021drag} achieved a maximal drag reduction of 23\% with opposition control in turbulent channel flow at Mach numbers $M_b = 1.5$.

Another promising approach involves the use of porous media. In compressible turbulent flows, porous surfaces are particularly effective for managing thermal loads, enhancing heat transfer in thermal protection systems, which makes them highly relevant for aerospace applications. These techniques offer significant potential for improving both aerodynamic performance and thermal management in high-speed compressible flows. Recent experimental and numerical studies have demonstrated that by tuning the permeability of the porous medium, it is possible to achieve significant modulation of both turbulence intensity and drag, providing a approach to drag reduction in practical applications \citep[]{kim2018experimental, manes2009turbulence, Rosti.2018, chu2021transport,Gomez.2019,Lacis.2020}. For instance, in thermal protection systems (TPS) used in high-speed aerospace applications \cite[]{mansour2024flow}, effective flow control can significantly reduce heat loads and improve material longevity, ensuring better thermal management and structural integrity. \citet{chen2021effects} studied flows in channels at Mach numbers of $M_b = 1.5$ and $M_b = 3.5$, where the channel walls were modeled using a time-domain impedance boundary condition \citep{chen2021effects,chen2021trapped}. LES were employed to examine how porous walls influence the flow, particularly with respect to pressure changes and stress distribution. These findings highlight the potential of porous media in advancing the design of efficient flow systems, offering enhanced control over turbulent structures and contributing to both aerodynamic performance and thermal regulation.

%Optimization methods have played a critical role in fine-tuning these control strategies \cite[]{wang2024opposition,sonoda2023reinforcement}. As flow control techniques grow in complexity, optimization approaches are needed to identify the best control parameters. Traditional optimization methods, such as adjoint-based gradient approaches \cite[]{bewley2001dns}, have been widely applied to flow control problems. These methods compute the gradients of an objective function, such as drag reduction or heat transfer enhancement, to guide the optimization process. 

While traditional flow optimization and control methods have demonstrated effectiveness in specific scenarios, they often rely on empirically tuned parameters. To address these challenges, the field has increasingly turned to advanced control and optimization techniques that provide a more systematic and rigorous approach. Among these, adjoint-based optimization has become a cornerstone in the pursuit of efficient flow control and design, offering a mathematically robust framework for optimizing performance across a wide range of flow configurations \cite[]{kungurtsev2019adjoint} or data assimilation \cite[]{plogmann2024adjointbasedassimilationsparsetimeaveraged}.
%Adjoint-based optimization has become a cornerstone in the field of flow control and design optimization due to its computational efficiency in handling high-dimensional problems. 
The adjoint method allows the efficient computation of gradients of an objective function, such as drag, lift, heat transfer, or energy dissipation, with respect to a large number of control variables by solving the adjoint equations derived from the governing Navier-Stokes equations. This approach is particularly advantageous because the cost of computing the gradient is independent of the number of design parameters, making it highly suitable for complex systems, such as aerodynamic shape optimization, where traditional optimization techniques would be prohibitively expensive \cite[]{jameson1988aerodynamic}. Adjoint-based techniques have been successfully applied to a wide range of flow control problems, including boundary layer manipulation, turbulence suppression, and noise reduction \citep{bewley2001dns,kim2003control}.

Despite its strengths, adjoint optimization faces several challenges. One of the primary difficulties lies in the derivation and implementation of the adjoint equations, particularly for unsteady, turbulent, or multiphase flows. For these complex flows, the adjoint equations must be accurately formulated and solved in tandem with the primal equations, leading to significant numerical complexity and computational expense. Additionally, the presence of discontinuities in the flow, such as shock waves or regions of flow separation, can lead to difficulties in ensuring stable and convergent adjoint solutions, as these sharp gradients are challenging to resolve in both the forward and adjoint simulations \cite[]{giles2000introduction}. Furthermore, adjoint-based methods are often limited by the need for accurate linearization of complex physical models, and the derivation of adjoint systems for industrial-scale solvers can be time-consuming and error-prone. Moreover, adjoint methods can struggle with handling non-smooth optimization landscapes, particularly in turbulent or chaotic flow regimes \cite[]{vishnampet2015practical}, where the adjoint variables exhibit high sensitivity to small changes in the control inputs, leading to slow convergence or even divergence in optimization. 

Machine learning (ML) has also emerged as a promising avenue for fluid mechanics \cite[]{Duraisamy.2019,vinuesa2022enhancing,srinivasan2019predictions,yang2024data,chu2024non,han2024combined}.
%Techniques such as Bayesian Optimization (BO) and Reinforcement Learning (RL) have been applied to explore control landscapes in a more adaptive and intelligent manner. 
Bayesian Optimization (BO), in particular, is suited for cases where objective function evaluations are expensive, as it uses probabilistic models to identify promising regions of the control space. This method has been used to optimize flow control strategies in scenarios where traditional gradient-based methods may struggle, offering a more global search that accounts for uncertainty in the control space.
\cite{mahfoze2019reducing} developed a BO framework to optimize low-amplitude wall-normal blowing control in a turbulent boundary-layer flow. The BO framework identifies the optimal blowing amplitude and coverage, achieving up to a 5\% net power savings within 20 optimization iterations, which require 20 DNS.
RL, on the other hand, has been increasingly explored for flow control applications where the control strategy can be learned through interaction with the flow environment \cite[]{sonoda2023reinforcement}. RL algorithms allow for agents to learn optimal control policies through trial and error. In the context of flow control, RL has demonstrated potential in complex, nonlinear flow configurations, where direct gradient methods may not provide effective solutions. 

In addition to adjoint methods and ML, differentiable fluid dynamics provides a novel approach that combines the strengths of traditional gradient-based methods with the flexibility of ML frameworks. In differentiable fluid dynamics, the entire fluid simulation becomes differentiable, allowing the efficient computation of control parameter gradients using automatic differentiation (AD) \cite[]{list2022learned,kochkov2021machine}. Automatic differentiation is implemented through computational graphs, which represent the sequence of mathematical operations executed during the forward pass. Each operation in the graph is a node, and the edges represent the flow of data between operations, allowing dependencies between variables to be tracked. In reverse-mode of AD, the graph is traversed backward after the forward pass, applying the chain rule to propagate gradients efficiently. This enables fast computation of gradients, even in large, deep models. This way, AD allows for the direct calculation of exact gradients of the objective function, eliminating the need for hand-derived adjoints or computationally expensive finite-difference approximations \cite[]{alhashim2024engineering}. 

Recently, differentiable fluid dynamics has gained significant traction due to its ability to seamlessly integrate with ML frameworks \cite[]{toshev2024jax,ataei2024xlb}, enabling the use of fast-evolving data-driven approaches in fluid simulations. Beyond just optimization, it holds broader potential in areas like data assimilation \cite[]{buhendwa2024shape} and data-driven modeling \cite[]{list2022learned,fan2024differentiable}, allowing for the incorporation of governing equations directly into the learning process. This approach bridges the gap between purely data-driven models and physics-based simulations, enabling more accurate and reliable modeling of complex, chaotic systems.

In this work, we propose an AD-based optimization framework for flow control in compressible turbulent channel flows. By leveraging the AD capability of differentiable fluid dynamics, we are able to compute the exact gradients of objective functions, allowing us to efficiently optimize control strategies involving opposition control or porous media. 
Through the application of AD in the framework of differentiable fluid dynamics, we show that the flow optimization and control can be performed in an easier and computationally efficient manner. This work highlights the potential of differentiable fluid dynamics for end-to-end optimization for complex flow control scenarios. \S\ref{sec:num} presents the numerical techniques employed in this study, covering the differentiable fluid dynamics framework as well as the optimization workflow. In \S\ref{sec:results}, the effectiveness of drag reduction through opposition control and permeable wall designs will be illustrated under varying optimization scenarios. \S\ref{sec:conclusion} provides a conclusive discussion.

\section{Numerical approach}\label{sec:num}

\subsection{Differentiable solver: JAX-Fluids}

In the present study, we establish our optimization framework utilizing JAX-Fluids \cite[]{bezgin2023jax,bezgin2024jax}, a Python-based CFD solver with full differentiability, tailored for compressible single- and two-phase flow scenarios. JAX-Fluids integrates high-order Godunov-type finite-volume methodologies with positivity-preserving limiters to enhance robustness. By leveraging JAX primitives, it facilitates efficient parallel processing on GPU and TPU platforms. This integration allows JAX-Fluids to perform direct numerical simulation (DNS) of complex flow dynamics with high-order precision and computational efficacy. 
A key feature of JAX-Fluids is its differentiability, enabled through JAX’s automatic differentiation (\texttt{autodiff}) capabilities. This allows for the computation of derivatives of scalar outputs (e.g., a loss function) with respect to any of the input parameters, such initial conditions and physical properties of the flow. Therefore, the solver is ideally suited for present optimization purposes due to its high computational efficiency and differentiability.

However, the current JAX-Fluids does not support AD on its built-in boundary conditions. To optimize the control parameters at the boundaries, we developed an extended boundary condition framework that allows for the complete differentiability of boundary conditions. This unlocks new possibilities for optimizing flow control strategies directly through gradient-based methods.
The primary idea is to construct a user-defined container that holds the boundary condition parameters and pass it from the high-level function, such as \texttt{feed\_forward()}, into the JAX-Fluids computational pipeline, and reach low-level halo cell update function, where boundary conditions are enforced. During each time step, the boundary parameters interact with the flow solver by being passed into the halo cell update function, where boundary conditions are applied. 

In JAX, functions that involve arrays or computational operations are traced to construct a computational graph. Since the boundary parameter array is part of the traced computation, it becomes part of this graph. Since the boundary parameters are passed through the same computational graph, their gradients can be computed during backward pass. This enables gradient-based optimization of these boundary parameters to minimize or optimize a loss function. The expanded patch allowing for differentiable boundary conditions in JAX-Fluids has been made publicly available (\href{https://github.com/WangWen-kang/JAX-BC.git}{https://github.com/WangWen-kang/JAX-BC.git})

%\subsection{Governing equations}
\subsection{Governing equations and numerical methods}

The optimization and control is based on the DNS of compressible turbulent channel flow, and the Navier-Stokes equations of conservative variables are,

\begin{equation}
\frac{\partial \mathbf{U}}{\partial t} + \frac{\partial \mathbf{F}^c(\mathbf{U})}{\partial x} + \frac{\partial \mathbf{G}^c(\mathbf{U})}{\partial y} + \frac{\partial \mathbf{H}^c(\mathbf{U})}{\partial z} = \frac{\partial \mathbf{F}^d(\mathbf{U})}{\partial x} + \frac{\partial \mathbf{G}^d(\mathbf{U})}{\partial y} + \frac{\partial \mathbf{H}^d(\mathbf{U})}{\partial z} + \sum_i \mathbf{S}_i(\mathbf{U})
\label{NSE}
\end{equation}
where $\mathbf{F}^c$, $\mathbf{G}^c$, and $\mathbf{H}^c$ denote the convective fluxes in $x$-, $y$- and $z$-direction. Analogously, $\mathbf{F}^d$, $\mathbf{G}^d$, and $\mathbf{H}^d$ denote the dissipative fluxes in the three spatial dimensions. The right-hand side is complemented by the sum of all source terms $\sum_i \mathbf{S}_i(\mathbf{U})$.
The primitive variables are the fluid density $\rho$, the velocity components $u$, $v$, and $w$ (in $x$-, $y$-, and $z$-direction, respectively), and the pressure $p$. %$\mathbf{u}=[u,v,w]^T$ is the velocity vector. 
$E=\rho e+\frac{1}{2}\rho \mathbf{u} \cdot \mathbf{u}$ denotes the total energy of the fluid. The vector of the conservative variables is given as

\begin{equation}
\mathbf{U} = \begin{bmatrix}
\rho \\
\rho u \\
\rho v \\
\rho w \\
E \\
\end{bmatrix},
\end{equation}
and the convective fluxes are

\begin{equation}
\mathbf{F}^c(\mathbf{U}) = \begin{bmatrix}
\rho u \\
\rho u^2 + p \\
\rho u v \\
\rho u w \\
u(E + p) \\
\end{bmatrix}, \quad
\mathbf{G}^c(\mathbf{U}) = \begin{bmatrix}
\rho v \\
\rho u v \\
\rho v^2 + p \\
\rho v w \\
v(E + p) \\
\end{bmatrix}, \quad
\mathbf{H}^c(\mathbf{U}) = \begin{bmatrix}
\rho w \\
\rho u w \\
\rho v w \\
\rho w^2 + p \\
w(E + p) \\
\end{bmatrix}.
\end{equation}
The dissipative fluxes describe viscous effects and heat conduction.

\begin{equation}
    \mathbf{F}^d(\mathbf{U}) = \begin{bmatrix}
0 \\
\tau_{xx} \\
\tau_{xy} \\
\tau_{xz} \\
\sum u_i \tau_{ix} - q_x \\
\end{bmatrix}, \quad
\mathbf{G}^d(\mathbf{U}) = \begin{bmatrix}
0 \\
\tau_{yx} \\
\tau_{yy} \\
\tau_{yz} \\
\sum u_i \tau_{iy} - q_y \\
\end{bmatrix}, \quad
\mathbf{H}^d(\mathbf{U}) = \begin{bmatrix}
0 \\
\tau_{zx} \\
\tau_{zy} \\
\tau_{zz} \\
\sum u_i \tau_{iz} - q_z \\
\end{bmatrix}
\end{equation}
The viscous stress is given by

\begin{equation}
\tau_{ij} = \mu \left( \frac{\partial u_i}{\partial x_j} + \frac{\partial u_j}{\partial x_i} - \frac{2}{3}\delta_{ij}\frac{\partial u_k}{\partial x_k} \right),
\end{equation}
where $\mu$ is the dynamic viscosity. The energy flux vector $\mathbf{q}= [q_x, q_y, q_z]^T$ is expressed via Fourier’s
heat conduction law, $\mathbf{q} = -\lambda \nabla T $, where $\lambda$ is the heat conductivity. 

The system of governing equations is closed by the ideal gas equation of state:

\begin{equation}
    p = (\gamma - 1)\rho e %- \gamma p_\infty
\end{equation}

\begin{equation}
    c = \sqrt{\frac{\gamma (p )}{\rho}}
\end{equation}
where ratio of specific heats \( \gamma = 1.4 \). In addition, we employ a simple power law model for the dynamic viscosity $ \mu $,

\begin{equation}
    \mu = \mu_{\text{ref}} \left( \frac{T}{T_{\text{ref}}} \right)^{0.7},
    \label{eq:mu}
\end{equation}
where \( \mu_{\text{ref}} \) is the dynamic viscosity at the reference temperature \( T_{\text{ref}} \). 
The thermal conductivity $\lambda$ is determined using a constant Prandtl number $Pr$ = 0.7. 

\begin{equation}
    \lambda = \frac{Pr}{\mu c_p},
\end{equation}
In this context, \( c_p \) represents the specific heat capacity at constant pressure.

The source terms \( \mathbf{S(U)} \) in Eq. \ref{NSE} represent body forces and heat sources. In current study, a constant mass flow rate is maintained by applying a body force in the streamwise (\(x\)) direction using a PID controller that minimizes the error between the target and current mass flow rate, 

\begin{equation}
    e(t) = \frac{\dot{m}_{\text{target}} - \dot{m}(t)}{\dot{m}_{\text{target}}}.
\end{equation}
 %For the calculation of the convective intercell numerical fluxes, WENO-type high-order discretization schemes in combination with approximate Riemann solvers is used. Dissipative fluxes are calculated by central finite-differences.  A third-order Runge–Kutta (RK3) time integration scheme is used to ensure stability and accuracy of the solution.Furthermore, the level-set implementation allows for arbitrary immersed solid boundaries.

\begin{figure}
    \centering
    \includegraphics[width=0.6\linewidth]{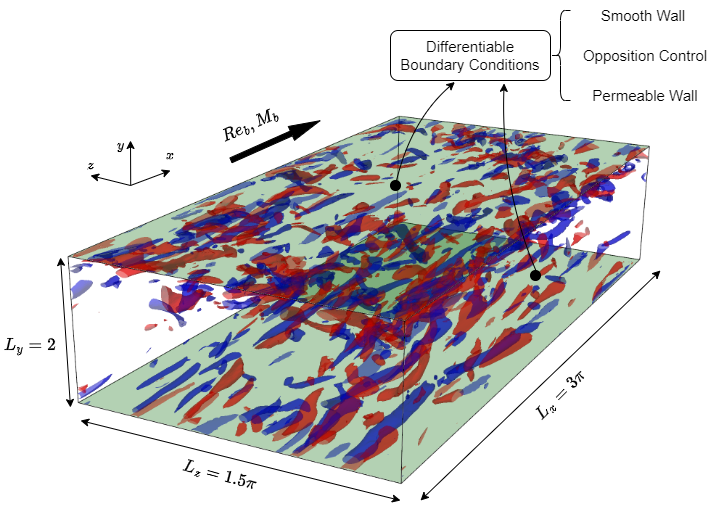}
    \caption{Domain of compressible DNS channel flow. The snapshot of the flow field, extracted from the smooth wall channel, serves as the initial condition for control. The blue and red isosurfaces represent streamwise vorticity at levels $\omega_x=\pm\sigma_x$.}
    \label{fig:DNS_channel}
\end{figure}

%$Ma_b=1.5$ and $Re_{\tau}=180$
%The domain size is $3\pi \times 2 \times 1.5$ ($x \times y \times z$) 
%$192\times128\times96$ ($x \times y \times z$) 
%isothermal boundary condition on both walls
The computational domain size ($L_x/h \times L_y/h \times L_z/h $) is \( 3\pi \times 2 \times 1.5\pi \) in the streamwise (\(x\)), wall-normal (\(y\)), and spanwise (\(z\)) directions, respectively (figure \ref{fig:DNS_channel}), where $h$ is channel half width. The grid resolution consists of \( 192 \times 128 \times 96 \) points in the corresponding \(x \times y \times z\) directions. 
The DNS grid is uniform in the streamwise (\(x\)) and spanwise (\(z\)) directions, while a hyperbolic-tangent stretching is applied in the wall-normal (\(y\)) direction with a stretching factor of  1.8. The grid resolution in the streamwise and spanwise directions is \(\Delta x^+ = \Delta z^+ = 10.71\) with $\Delta x^+=xu_\tau/\nu$, $\Delta z^+=zu_\tau/\nu$ and $ u_\tau = \sqrt{\tau_w / \rho} $. The cell sizes in the wall-normal direction vary, with a minimum value of \(\Delta y^+_{\text{min}} = 0.69\) and a maximum value of \(\Delta y^+_{\text{max}} = 6.48\).
A validation of the present DNS against the DNS data from \citet{yao2021drag} is provided in Appendix \ref{app:mean}.
%The spatial coordinates are nondimensionalized by the channel half-width, with velocity scaled by the friction velocity \( u_\tau = \sqrt{\tau_w / \rho} \) and temperature scaled by the wall temperature \( T_w \).

The bulk density is computed as $\rho_b = \frac{1}{2h} \int_{-h}^{h} \langle \rho \rangle \mathrm{~d}y$, and the bulk velocity is calculated as \( U_b = \frac{1}{2h \rho_b} \int_{-h}^{h} \langle \rho u \rangle \mathrm{~d}y \).
The Reynolds number, based on bulk density, bulk velocity, channel half-width, and wall viscosity, is $Re_b = \frac{\rho_b U_b h}{\mu_w} = 3000$.
The bulk Mach number, defined as the ratio of the bulk velocity to the speed of sound at the wall, is given by \( Ma_b = \frac{U_b}{c_w} = 1.5 \).
Isothermal no-slip boundary conditions are enforced at the channel walls, where \( T = 1 \) and \( u = 0 \) at \( y = \pm h \). Periodic boundary conditions are applied in both the streamwise and spanwise directions. 

For the numerical setup, we employ a TENO6-A \cite[]{fu2016family} cell-face reconstruction method combined with an HLLC Riemann solver. The TENO6-A reconstruction is enhanced by an interpolation limiter, while flux limiters are not utilized in this work since the single-phase cases considered do not involve strong shock discontinuities. Diffusive fluxes are discretized using sixth-order central finite-difference schemes, and the temporal evolution is carried out using a third-order TVD Runge-Kutta (TVD-RK3) scheme with a CFL number of 0.9.

\subsection{The flow control boundary conditions}

%Opposition control BC:  $u=w=0, v\ne 0$, net flux as zero, constant in time
In addition to the smooth wall channel flow, which taken as the baseline of flow control performance, we investigate two types of flow control boundary conditions. The first is the boundary condition of opposition control. A wall normal velocity 
\begin{equation}
    \bm{u}=-\phi(\bm{x},t)\cdot\bm{n}
\end{equation}
is applied on the upper and lower wall, $\Gamma^+$ and $\Gamma^-$, where $\bm{n}$ is the unit outward normal to the boundary. This control strategy is applied with the objective of introducing a counteracting wall-normal velocity at the boundary, designed to oppose the near-wall turbulence structures. The total net flux across the boundary is constrained to be zero,
\begin{equation}
    \int_{\Gamma^+} \phi \mathrm{~d}\bm{x}=\int_{\Gamma^-} \phi \mathrm{~d}\bm{x}=0,
\end{equation}
 ensuring that there is no net mass flow through the wall over time.

%Porous media BC: $u=w=0, v=-\beta p^{\prime}$ \cite{Jimenez.2001}, $\beta$ is constant and has a range of $0-0.7$ according to \cite[]{Jimenez.2001}.

The second boundary condition is a permeable wall boundary condition proposed by \citet{Jimenez.2001}. The wall normal velocity on the lower and upper walls are modeled as 
\begin{equation}
     \bm{u} = -\beta(\bm{x},t) p'\cdot\bm{n} 
\end{equation}
where the parameter \( \beta \) works like the permeability of the wall, and modulates the coupling between the wall-normal velocity and the pressure fluctuations. The value of \( \beta \) varies in the range of 0 to 0.7, as suggested by \citet{Jimenez.2001}. In contrast to the initial research by \citet{Jimenez.2001}, where $\beta$ remains constant across space and time, our approach considers $\beta$ as variable in both dimensions in order to enhance its control capabilities.

Note that the isothermal condition ($T=1$) is applied at the channel walls for both the opposition control and permeable wall boundary conditions. This is consistent with the baseline case with smooth wall channel.

\subsection{Optimization workflow}

%Optimizer, Adam \cite[]{kingma2014adam}
%Time advancing \cite{bewley2001dns}
%$t^+=50/100$ for one optimization horizon.
%$t^+=2000$ corresponds to 20 flow through time.

\begin{figure}
    \centering
    \includegraphics[width=1.0\linewidth]{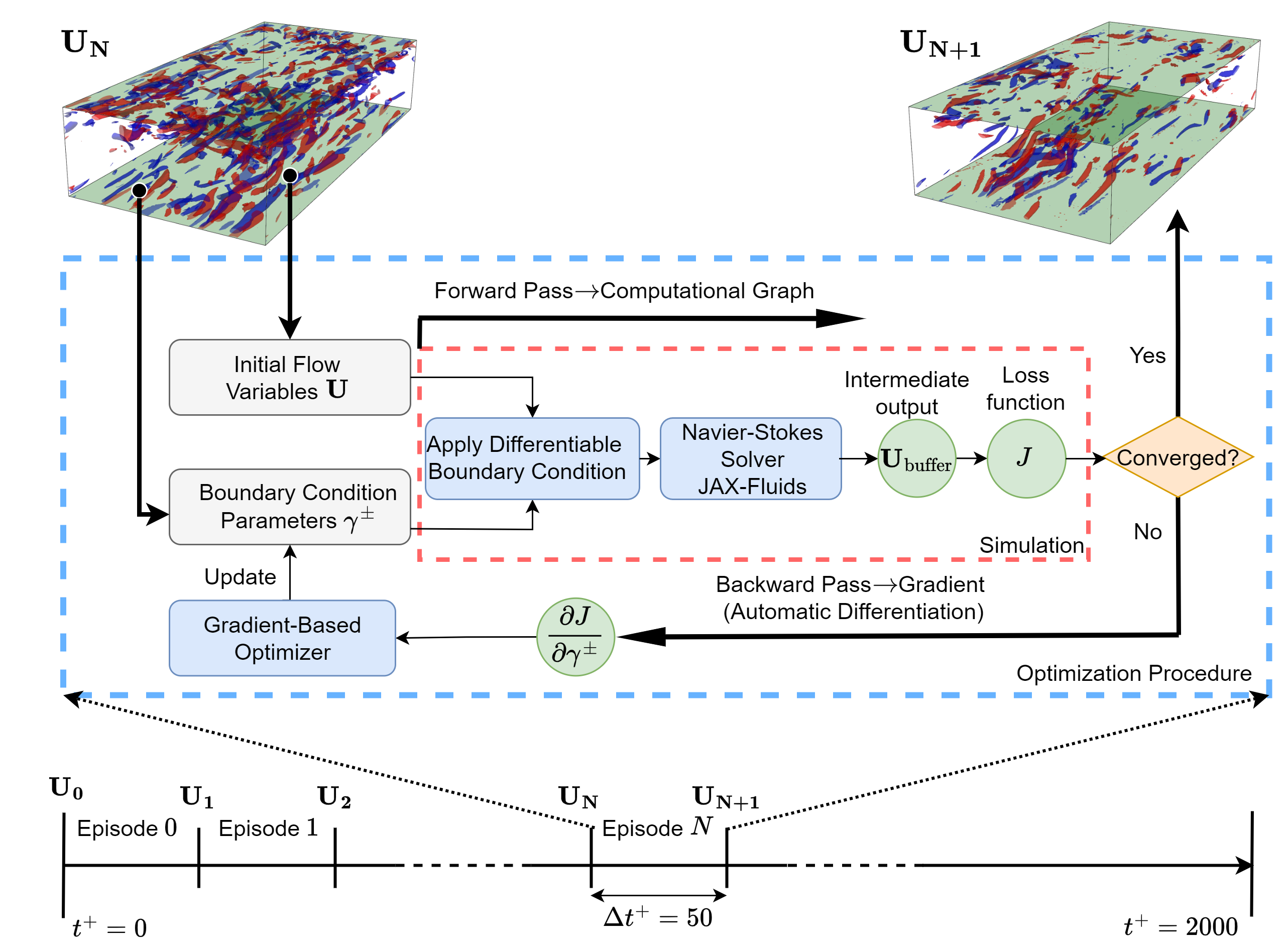}
    \caption{The procedure for automatic differentiation and flow control optimization.}
    \label{fig:AD}
\end{figure}

Figure \ref{fig:AD} illustrates the workflow for the AD-based optimization.  We adopted the receding-horizon predictive control process introduced by \cite{bewley2001dns}. The evolution of the flow consists of a series of `episodes'. In each episode, the optimized control parameters is explored by an iterative gradient based optimization algorithm. Once this iteration converges, the flow is advanced to the next episode, and the optimization process is initiated again.

Consider episode $N$ as an instance. The optimization's input is $\bf{U}_N$, representing the terminal state from episode $N-1$. Alongside the boundary condition parameters for the upper and lower walls, $\gamma^\pm$, the initial flow variables are processed through the differentiable solver to yield an intermediate output $\bf{U}_{\mathrm{buffer}}$, which is used to calculate the loss function $J$. During the forward computation, the computational graph is created, allowing for the calculation of the gradient of the loss function concerning the boundary parameter $\frac{\partial J}{\partial \gamma^\pm}$ via AD. This requires just one computation pass to form the graph. The gradient-based optimizer then updates control inputs for the differentiable boundary conditions. In this context, the Adam optimizer \cite[]{kingma2014adam} is employed, minimizing the loss function using the derived gradients. When optimization converges, the boundary condition parameters are fixed for the current episode, enabling the flow to proceed by $\Delta t$ and generate the terminal flow state $\bf{U}_{N+1}$.
 
For the current channel domain, the upper and lower walls consist of $192\times96\times2$ grid points, resulting in a 36,864 dimensional optimization problem. AD is particularly powerful with functions with such high dimensional input. Only one backward pass through the computational graph is needed to compute the gradient with respect to all input variables, since each partial derivative is accumulated in parallel during the backward pass. Therefore, the total cost remains low even as the input dimension increases. In addition, JAX uses the XLA (Accelerated Linear Algebra) compiler for just-in-time (JIT) compilation of functions. JIT compilation translates the Python functions into a lower-level, optimized representation that can run directly on the hardware, avoiding the overhead of Python's interpreter. This is particularly beneficial for AD, where operations need to be traced through a computational graph. With JIT, this tracing happens once, and the compiled version can be executed repeatedly without needing to re-trace.

The duration of the optimization episodes plays a vital role in the system. This time horizon influences how far ahead the control algorithm projects the system's behavior. A longer horizon enables the controller to foresee long-term effects of its actions, though it demands more computational power. Conversely, a shorter horizon decreases computational effort and favors real-time execution, but might lead to less optimal decisions. In our current study, we evaluate two time horizons: $\Delta t^+=\Delta t u^2_\tau/\nu_w=25$ and $50$, with the latter corresponding to 0.69 flow-through time $tU_b/L_x$. 
The total simulation runs for \( t^+ = 2000 \), translating to 27.6 flow-through times, which is adequate for the controlling boundary to significantly affect the turbulent flow dynamics.
The simulation and optimization processes utilize 8 NVIDIA A100 GPUs on a single node within the HAWK-AI infrastructure at the High Performance Computing Center Stuttgart (HRLS). Each optimization with \( t^+ = 2000 \) takes about two days of wall-clock time. This duration is manageable for industrial applications, highlighting its potential in practical optimization tasks. Additionally, the approach can be adapted for larger-scale problems by incorporating more GPU resources.

\subsection{Loss functions}\label{sec:loss_function}
The choice of loss function has a large impact on the optimization process. In the current study, we compare several types of loss functions.

\subsubsection{Cumulative wall friction control}
To consider the cumulative effect of wall friction $\tau_w=\mu\left(\frac{\partial u}{\partial y}\right)_{y=0}$, the loss function can be formulated as the integration of friction drag on the upper and lower walls and over the time horizon (0,$\Delta t$). For the opposition control strategy, the magnitude of $\phi$ also needs to be constrained to limit the cost of control, i.e.,

\begin{equation}
J_{\tau(\mathrm{cum})}(\phi)=-\frac{1}{\Delta t}\int_0^{\Delta t} \int_{\Gamma^{ \pm}} \mu \frac{\partial u(\phi)}{\partial n} \mathrm{~d} \boldsymbol{x} \mathrm{~d} t+\frac{\ell^2}{2} \int_{\Gamma^{ \pm}}\rho_w \phi^2 \mathrm{~d} \boldsymbol{x} .
\end{equation}
where $\partial / \partial n$ represents the gradient in the direction perpendicular to the wall, facing outward. The factor $\ell^2$ represents the price of control, which regulates the importance of the control cost in the loss function, and hence affects the optimization result of $\phi$. In the current work, we choose $\ell^2$ to be 1. As will be shown later, this is a relatively large constraint on the input energy of opposition control and results in a small amplitude of $\phi$. 

For permeable wall cases, there is no energy consumption except at the state-changing instants between control sections. In this study, we ideally assume that the state-changing process of permeable walls has a marginal cost, hence the cost function for permeable wall cases is

\begin{equation}
J_{\tau(\mathrm{cum})}(\beta)=-\frac{1}{\Delta t}\int_0^{\Delta t} \int_{\Gamma^{ \pm}} \mu \frac{\partial u(\beta)}{\partial n} \mathrm{~d} \boldsymbol{x} \mathrm{~d} t.
\end{equation}

\subsubsection{Cumulative turbulent kinetic energy control}

In the present study, the longest time horizon for each episode, $\Delta t^+ = 50$, corresponds to 0.69 of the flow-through time. This is similar to the time horizon employed in earlier research \cite[]{bewley2001dns}, yet it remains insufficient for the entire channel to fully stabilize following the implementation of control. Therefore, taking the quantity of interest directly as the cost function may not be the most effective and stable means of reducing it over the long term. In particular, wall friction only involves the information on the wall, which can be manipulated easily by setting the boundary condition in a short period. The potential long-term effects of these manipulations on the outer region are not taken into account, which may lead to instability of the control results.

Turbulence in the near-wall region induces wall-normal convective transport, thereby enhancing both drag and heat transfer in the flow. It is well known that turbulent production throughout the channel significantly contributes to wall friction \cite[]{Renard_Deck_2016}. Hence, alleviating turbulence intensity might lead to a reduction in wall friction. Moreover, TKE, in contrast to wall friction which is concentrated in the region close to the wall, results from the turbulence sustaining processes occurring throughout the entire channel. Consequently, TKE may serve as a more effective loss function than wall friction in producing stable control outcomes. We employed the cumulative TKE across the channel as the loss function. For the opposition control boundary, we also include the constraint on the cost of control. 

\begin{equation}
    J_{k(\mathrm{cum})}(\phi)=\frac{1}{4h\Delta t} \int_0^{\Delta t} \int_{\Omega}
    \rho|\boldsymbol{u}(\phi)|^2 \mathrm{~d} \boldsymbol{x} \mathrm{~d} t+\frac{\ell^2}{2} \int_{\Gamma^{ \pm}}\rho_w \phi^2 \mathrm{~d} \boldsymbol{x}.
\end{equation}
For the permeable wall case, the cost of control neglected hence the loss function is
\begin{equation}
    J_{k(\mathrm{cum})}(\beta)=\frac{1}{4h\Delta t} \int_0^{\Delta t} \int_{\Omega}
    \rho|\boldsymbol{u}(\beta)|^2 \mathrm{~d} \boldsymbol{x} \mathrm{~d} t.
\end{equation}

\subsubsection{Terminal wall friction and TKE control}
In addition to cumulative control that considering wall friction or TKE throughout the time horizon, another control method is to concentrate solely on the terminal value of each episode. Using the terminal control approach, the cost functional does not penalize deviations in the quantity of interest during the intermediate stages of each epsiode, provided that these deviations result in lower values of the quantity of interest at the end of each optimization horizon. Compared to traditional methods, the terminal control approach offers greater flexibility in control strategies and potentially better control outcomes; however, choosing an inappropriate loss function and having a very short time horizon may also result in instability.

For terminal  wall friction minimization with opposition control, the loss function is

\begin{equation}
    J_{\tau(\mathrm{ter})}(\phi)=-\int_{\Gamma^{ \pm}} \mu \frac{\partial u(\phi;\Delta t)}{\partial n} \mathrm{~d} \boldsymbol{x}+\frac{\ell^2}{2} \int_{\Gamma^{ \pm}}\rho_w \phi^2 \mathrm{~d} \boldsymbol{x},
\end{equation}
and the loss function for permeable wall is

\begin{equation}
    J_{\tau(\mathrm{ter})}(\beta)=-\int_{\Gamma^{ \pm}} \mu \frac{\partial u(\beta;\Delta t)}{\partial n} \mathrm{~d} \boldsymbol{x}.
\end{equation}

Similarly, the loss functions for terminal TKE minimization are 

\begin{equation}
    J_{k(\mathrm{ter})}(\phi)=\frac{1}{4h} \int_{\Omega}
    \rho|\boldsymbol{u}(\phi;\Delta t)|^2 \mathrm{~d} \boldsymbol{x}+\frac{\ell^2}{2} \int_{\Gamma^{ \pm}}\rho_w \phi^2 \mathrm{~d} \boldsymbol{x}    
\end{equation}
and 

\begin{equation}
    J_{k(\mathrm{ter})}(\beta)=\frac{1}{4h} \int_{\Omega}
    \rho|\boldsymbol{u}(\beta;\Delta t)|^2 \mathrm{~d} \boldsymbol{x}  
\end{equation}
for opposition control and permeable wall, respectively.

\section{Results}\label{sec:results}
In the current section, we show the simulation with optimized opposition control and permeable wall configurations.

\begin{figure}
    \centering
    \includegraphics[width=1\linewidth]{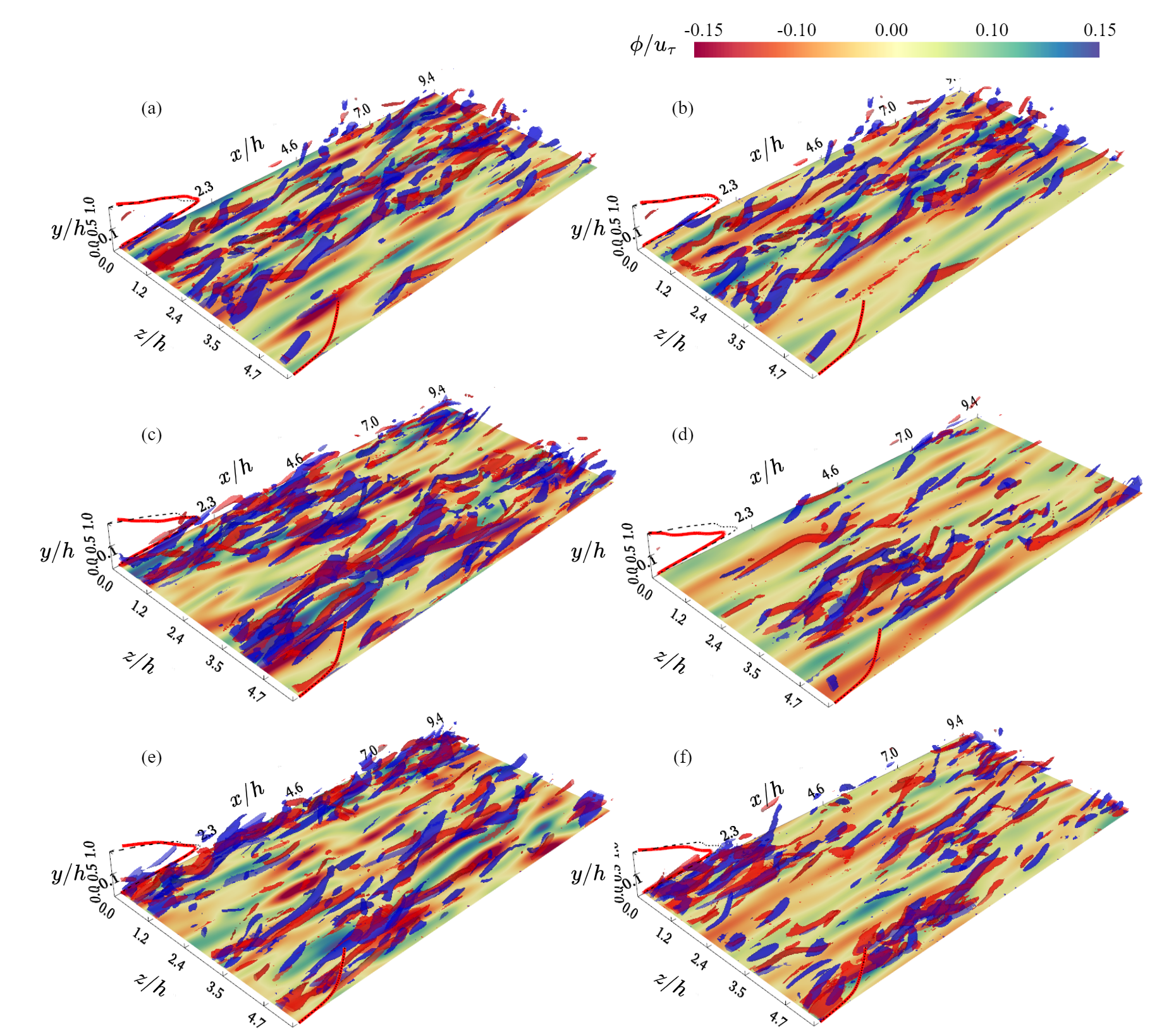}
    \caption{Vorticity structures above opposition control surfaces. Panels (a,c,e) show cumulative wall friction control with $\Delta t^+=50$, while panels (b,d,f) depict terminal TKE control with the same $\Delta t^+$. The pairs of panels (a,b), (c,d), and (e,f) represent $t^+=50$, $t^+=500$, and $t^+=2000$ from the onset of control, respectively. Blue and red iso-surfaces indicate streamwise vorticity at $\omega_x=\pm\sigma_{\omega_x}$, $\sigma_{\omega_x}$ being the standard deviation of $\omega_x$ at $t=0$. Control $\phi$ is illustrated on the wall with colored contours. Present snapshot profiles of $\langle u\rangle$ and $-\langle u'v'\rangle$ are overlaid on front ($z=1.5\pi$) and back ($z=0$) planes with solid red lines. Black dashed lines show profiles from the smooth wall scenario for comparison. See also supplementary videos 1 and 2 for the full simulation duration. }
    \label{fig:oppo_flow}
\end{figure}

\subsection{The performance of opposition control}
The performance of opposition control is determined by the control $\phi(t)$, which is influenced by the configuration of the optimization process. Since we use the auto-differentiation gradient as the optimization input, the form of the computational graph is the core. For the current setup, there are two important factors. First, the time horizon of the episodes decides the time integration length of the N-S equation, hence closely related to the complexity of the computational graph. It also defines the maximum flow field information available from the temporal dimension that could be utilized for optimization. 
Physically, the time horizon is the period during which the flow is allowed to develop under the same conditions $\phi$.  
Second, the choice of loss function also profoundly affects the computational graph, since it determines more specifically the variables, as well as the spatial and temporal range, included in the computational graph. Figure \ref{fig:oppo_flow} shows the comparison of instantaneous flow fields with different optimization targets.  

Initially, the vortices exhibit a similar shape and count in both scenarios at the early stage of control ($t^+=50$). However, as the flow continues to evolve ($t^+=500$ and 2000), the vortex structures diverge markedly in form and quantity between the two cases. In the cumulative wall friction control scenario (figure \ref{fig:oppo_flow} a,c,e), the vortices become more intense, whereas in the TKE control scenario (figure \ref{fig:oppo_flow} b,d,f), the turbulent structures are notably diminished. This divergence is evident in the mean statistics, such as the Reynolds stress profile $\langle u'v'\rangle$ (indicated by red solid lines on the plane $z=0$). Compared to the initial $\langle u'v'\rangle$ (denoted by black dashed lines on the plane $z=0$), the wall friction control case shows a slight increase in the magnitude of Reynolds stress, while the TKE control case registers a significant reduction in the peak of Reynolds stress. 

The rise in Reynolds stress in the case of wall friction control appears counter intuitive as Reynolds stress significantly contributes to friction drag. This is due to the modification of the mean velocity profile $\langle u\rangle$ by the control. However, this alteration is localized near the wall and is not distinctly visible (the $\langle u\rangle$ profiles are depicted as red solid lines in the plane $z=1.5\pi$ in figure \ref{fig:oppo_flow}).

In the following sections, we will assess the impact of control strategies across varying time horizons and loss functions. We will elucidate the drag reduction mechanisms for different scenarios in more detail.

\subsubsection{The development of wall friction under opposition control}
\begin{figure}
    \centering
    \includegraphics[width=0.8\linewidth]{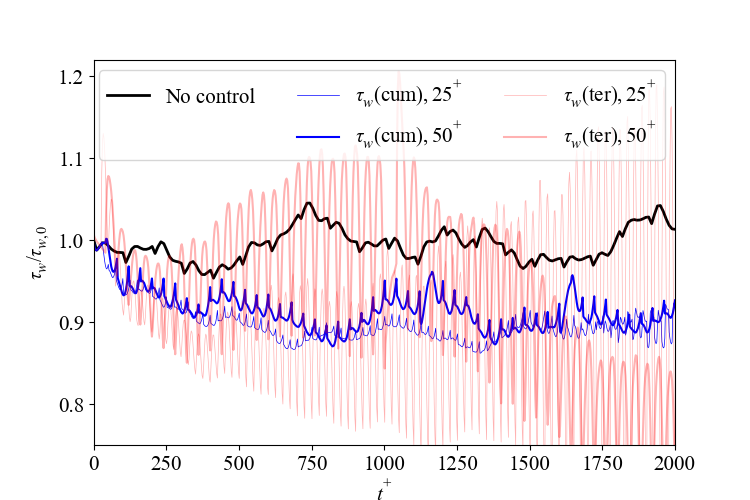}
    \caption{ The development of wall friction $\tau_w$ with opposition control directly targeting loss functions associated with $\tau_w$. In the legend, `$\tau_w$(cum), $25^+$' represents cumulative wall friction control with a time horizon of $\Delta t^+=25$. `$\tau_w$(ter)' indicates terminal wall friction control. This convention is consistently applied throughout the rest of the legend and the paper.}
    \label{fig:opposition-cf}
\end{figure}

In  \S\ref{sec:loss_function}, we introduced two types of loss functions directly targeting wall friction: the first, $J_{\tau(\mathrm{cum})}$, takes into account wall friction over the entire time horizon, while the second, $J_{\tau(\mathrm{ter})}$, focuses solely on the terminal value for each episode. Figure \ref{fig:opposition-cf} presents a comparison of the drag evolution history using both types of loss functions, analyzed over two different time horizon durations, $\Delta t^+=25$ and $50$.

For the cumulative $\tau_w$ control, the wall friction reduces to approximately 90\% of its initial value by $t^+=750$. A slight improvement in control outcomes is observed with a shorter time horizon of $\Delta t^+=25$. In contrast, when employing a terminal loss function $J_{\tau(\mathrm{ter})}$ and a time horizon of $\Delta t^+=25$, wall friction decreases by approximately 13\% at $t^+=400$, which seems to be more effective than cumulative control. However, after $t^+=500$, the performance deteriorates, exhibiting a significant fluctuation as well as a slow recovery in wall friction. This issue is exacerbated when terminal control is paired with an extended time horizon $\Delta t^+=50$, as the wall friction experiences strong fluctuations and fails to stabilize within the tested interval. 

The unstable performance of terminal control aligns with our expectations. As described in \S\ref{sec:loss_function}, focusing solely on the terminal value can lead to situations where only terminal friction is decreased. This is particularly true when combined with a wall friction-based loss function, as only the data points right above the wall is incorporated into the computational graph. In such cases, much of the channel flow dynamics is curtailed, and this truncation increases with longer time horizons. By focusing on the near-wall region and neglecting the rest of the turbulent channel, terminal control can become more aggressive and effective in the short term. However, this short-sighted approach may cause excessive disturbances, ultimately increasing turbulent fluctuations over time. To counteract the rise in wall friction from additional turbulence, terminal control requires more intense intervention, creating a vicious cycle, and resulting in control instability. The cumulative loss function, which considers the full evolution history of wall friction, offers an advantage by implicitly involving the dynamics of the outer region in the computational graph. Consequently, while cumulative control may take longer to demonstrate its effects, it tends to be more stable.

An alternative method to reduce wall friction involves suppressing the turbulence intensity within the channel, which is also the fundamental principle behind opposition control \cite[]{Choi_Moin_Kim_1994}. We also compare the performance of the cumulative and terminal type of loss function, i.e. $J_{k{(\mathrm{cum})}}$ and $J_{k{(\mathrm{ter})}}$, under time horizons $\Delta t^+=25$ and $\Delta t^+=50$. 

\begin{figure}
    \centering
    \includegraphics[width=0.8\linewidth]{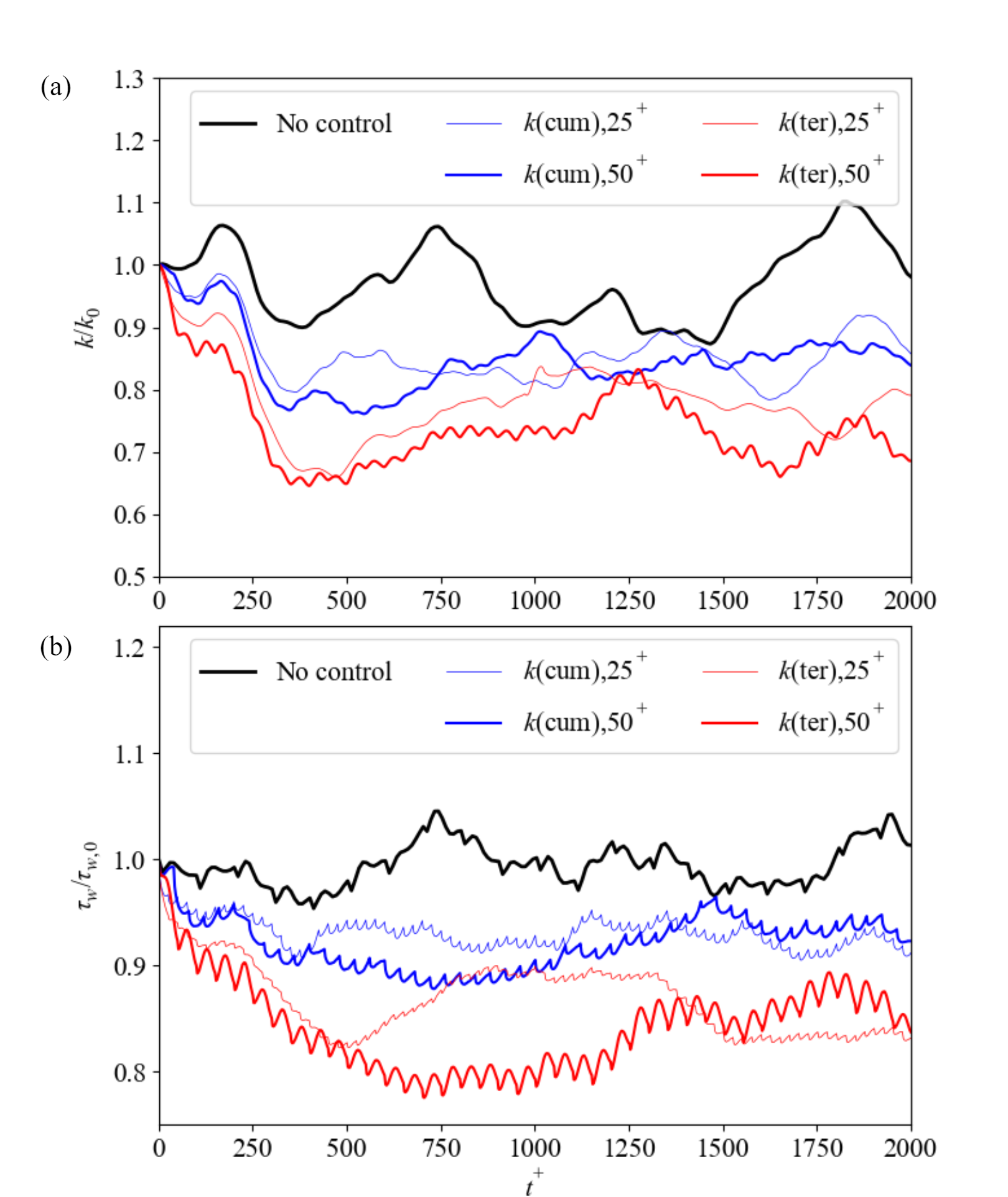}
    \caption{The history of (a) TKE $k$ and (b) wall friction $\tau_w$ with opposition control targeting TKE related loss function. The legend follows the same format as figure \ref{fig:opposition-cf}, with $k$ denoting TKE control. }
    \label{fig:opposition-k}
\end{figure}

Figure \ref{fig:opposition-k}(a) depicts the TKE evolution following the implementation of opposition control. With cumulative controls, TKE decreases by around 20\% at $t^+=250$. Initially, TKE declines slightly quicker with an extended time horizon, yet the difference between $\Delta t^+=25$ and 50 becomes insignificant after $t^+=750$. Terminal controls seem more efficient, lowering TKE by approximately 34\% at $t^+=250$. An extended time horizon seems advantageous for enhancing control performance. Despite a gradual recovery after the initial quick drop, the turbulence intensity remains significantly lower compared to the cumulative control scenarios.

In relation to the reduction of TKE, the wall friction in the TKE control scenarios also significantly declines (see figure \ref{fig:opposition-k}b). The cumulative control methods reduce wall friction by approximately 10\%, whereas the terminal control methods achieve a reduction of up to 20\%. Notably, the terminal control approach with a long time horizon of $\Delta t^+=50$ consistently surpasses the other configurations, demonstrating the most efficient wall friction control.

Unlike scenarios with wall friction control, where the terminal loss function yields unstable outcomes, controlling terminal TKE proves to be stable and significantly more effective. This aligns with findings by \cite[]{bewley2001dns}, who utilized optimization through the adjoint method. The key distinction between terminal friction and terminal TKE lies in the scope: TKE involves integrating turbulence intensity throughout the whole channel, whereas wall friction is concerned solely with data from the near-wall area. Wall friction can be quickly altered by adjusting boundary conditions, such as inducing slip velocity at the wall, but TKE cannot be significantly decreased in a brief time by merely altering boundary conditions, as it is tied to the processes of turbulence production, transport, and dissipation throughout the channel. Thus, while the terminal TKE loss function omits the intermediate values, the full flow dynamics of the channel is essential in the computational graph for its calculation. This ensures stability in the control from the ground up. Regarding its effectiveness, omitting intermediate values allows the opposition control to implement more aggressive control $\phi$ with greater flexibility. Further clarification is provided in the following sections.        

\subsubsection{The characteristic of control $\phi$}

\begin{figure}
    \centering
    \includegraphics[width=1\textwidth]{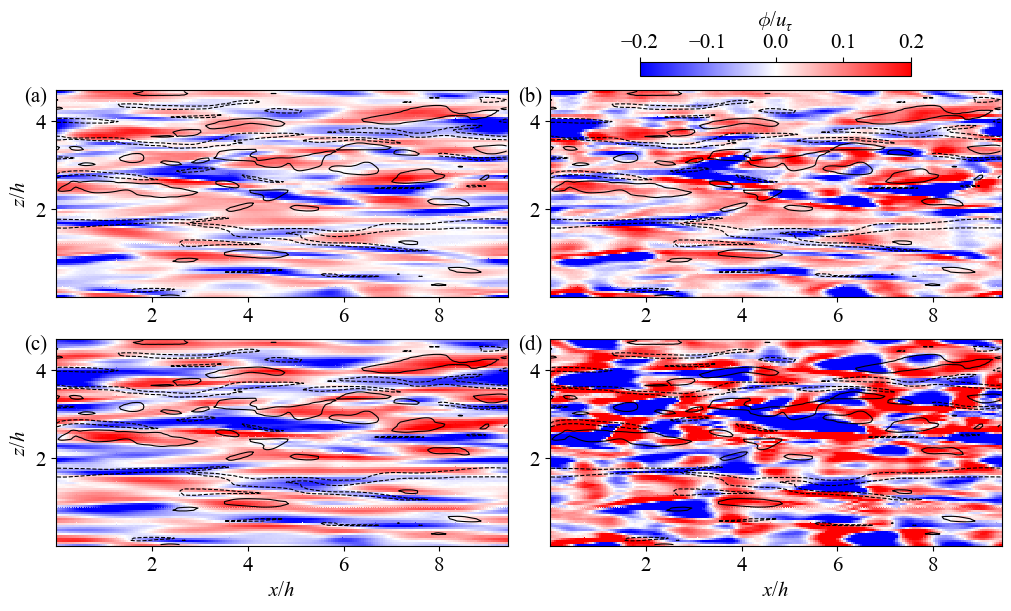}
    \caption{The opposition control $\phi(t^+=0)$ for wall friction control with different targets and time horizons. (a) cumulative $\tau_w$ with $\Delta t^+=25$; (b) terminal $\tau_w$ with $\Delta t^+=25$; (c) cumulative $\tau_w$ with $\Delta t^+=50$;(d) terminal $\tau_w$ with $\Delta t^+=50$. The isolines depict the initial $u'$ of the episode at buffer layer ($y^+=15$). The solid and dashed isolines indicate levels $u'/\sigma_{u'}=-1$ and 1, respectively.}
    \label{fig:phi_cf}
\end{figure}

\begin{figure}
    \centering
    \includegraphics[width=1\textwidth]{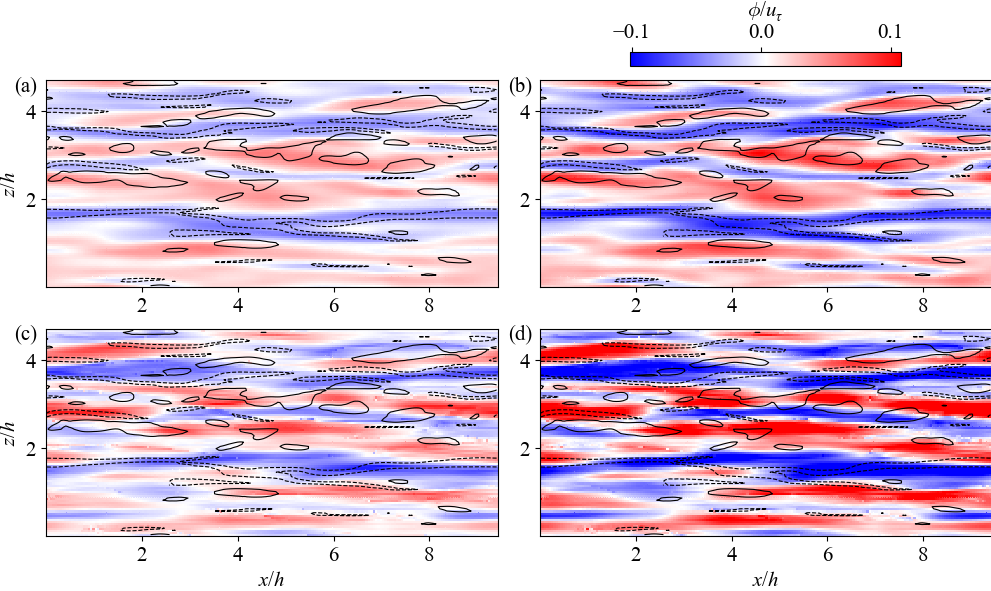}
    \caption{The opposition control $\phi(t^+=0)$ for TKE control with different targets and time horizons. (a) cumulative $k$ with $\Delta t^+=25$; (b) terminal $k$ with $\Delta t^+=25$; (c) cumulative $k$ with $\Delta t^+=50$;(d) terminal $k$ with $\Delta t^+=50$. The isolines depict the initial $u'$ of the episode at buffer layer ($y^+=15$). The solid and dashed isolines indicate levels $u'/\sigma_{u'}=-1$ and 1, respectively.}
    \label{fig:k_phi}
\end{figure}

To gain deeper insights into the control mechanisms and the differences arising from varying control targets, we examine in more detail the optimized $\phi$ across all opposition control scenarios. Figure \ref{fig:phi_cf} provides a comparison of $\phi$ fields in the wall friction control scenarios initialized under identical conditions. In the cases of cumulative wall friction control depicted in figure \ref{fig:phi_cf}a,c, the control fields $\phi$ exhibit streamwise elongated structures, resembling the form of high and low-speed streaks in the near-wall region (illustrated with isolines). When considering a longer time horizon, the control $\phi$ shows longer streamwise structures. For terminal control $\phi$, the spanwise spacing of the streak-like formations matches that of the cumulative control $\phi$, but these formations possess a shorter streamwise scale and greater magnitude.

The similarity of $\phi$ and the $u'$ structures suggests that the form of $\phi$ is intimately linked to the reduction of energetic structures near the wall. Additionally, for cumulative control $\phi$, regions of upward momentum flux largely coincide with high-speed streaks (solid isolines in figure \ref{fig:phi_cf}), whereas the areas of downward flux align with low-speed streaks (dashed isolines). This correlation generates positive Reynolds stress $\langle u'v'\rangle$ which acts against the original Reynolds stress in the flow. It is important to note that the $u'$ fields depicted with isolines in figure \ref{fig:phi_cf} represent the initial state of the optimization episode. There is no clear correlation between the terminal control $\phi$ and the initial $u'$ field because the terminal control is focused solely on the final outcome of the optimization. Since near-wall friction velocity can be quickly adjusted by altering the boundary condition, the influence from the initial flow state to the terminal control $\phi$ is anticipated to be minimal. It will be demonstrated subsequently that the terminal control $\phi$ displays a strong correlation with the terminal $u'$ fields. 

When compared to wall friction control, the TKE control fields $\phi$ display significant consistency across various loss functions (refer to figure \ref{fig:k_phi}). In all scenarios, the wall-normal flux in $\phi$ aligns closely with low and high-speed streaks within the buffer layer (indicated by isolines). Specifically, the upward flux corresponds to regions of positive $u'$, whereas the downward flux is situated in areas of negative $u'$. The impact of the duration of the time horizon is minimal. In cases with a longer time horizon, the coherent structures in control $\phi$ are smoothed due to time-averaging effects, resulting in fewer small-scale details. Terminal control $\phi$ exhibits a greater magnitude compared to cumulative control $\phi$, indicating a more aggressive control strategy. 

Terminal TKE control's similarity in control $\phi$ to cumulative TKE control supports our earlier discussion, indicating that loss functions involving terminal TKE implicitly encompass flow dynamics over the full time span. Consequently, even though the initial flow state isn't directly included in the loss function, the control $\phi$ closely resembles the energetic structures present within it.

\begin{figure}
    \centering
    \includegraphics[width=1\linewidth]{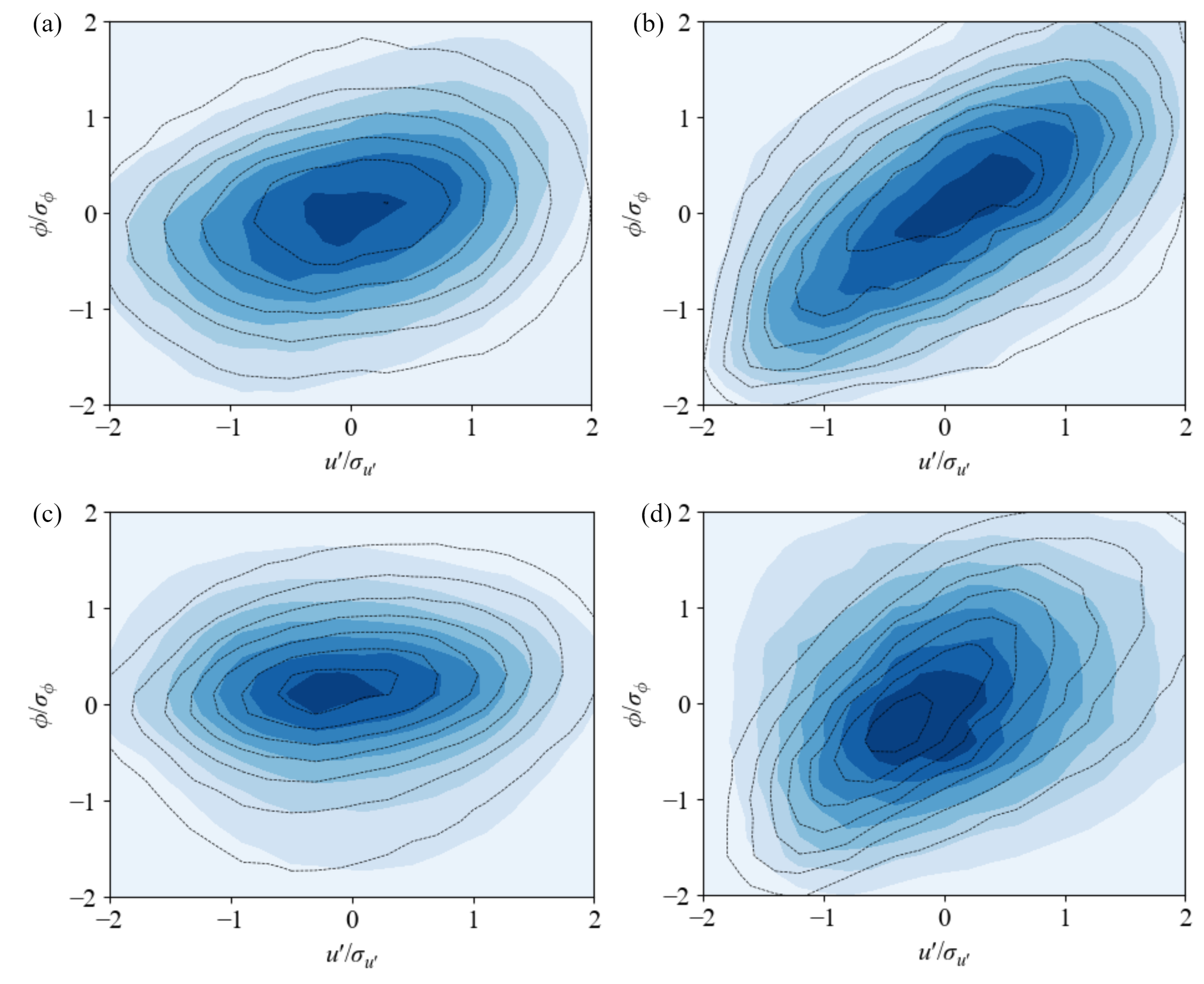}
    \caption{The joint probability density function (PDF) $f_{\phi u'}$ between control $\phi$ and streamwise fluctuation $u'$. (a) cumulative $\tau_w$ control with $\Delta t^+=50$; (b) cumulative $k$ control with $\Delta t^+=50$;(c) terminal $\tau_w$ control with $\Delta t^+=50$; (d) terminal $k$ control with $\Delta t^+=50$. The colored contour show the joint PDF $f_{\phi u'_0}$ between the control $\phi$ and initial $u'$, and the dashed isolines show the joint PDF $f_{\phi u'_{\Delta t}}$ between the control $\phi$ and terminal $u'$.}
    \label{fig:oppo_phiu}
\end{figure}

To gain a deeper insight into the regulation mechanisms for different control objectives, figure \ref{fig:oppo_phiu} displays the joint probability density function (PDF) relating the control variable $\phi$ to $u'$, highlighting their statistical association. For clarity, we illustrate only four typical scenarios with $\Delta t^+=50$, as other cases with shorter intervals exhibit essentially the same pattern. In cumulative wall friction control (with $\Delta t^+=50$, figure \ref{fig:oppo_phiu}a), the joint PDF $f_{\phi u'_0}$ (colored contour) between $\phi$ and the initial streamwise fluctuation $u'_0$ shows a distribution skewed towards the 1st and 3rd quadrants, indicating a clear positive correlation. This aligns with our earlier findings in instantaneous flow fields (figure \ref{fig:phi_cf}c). The joint PDF $f_{\phi u'_{\Delta t}}$ (dashed isolines) between $\phi$ and the final $u'_{\Delta t}$ exhibits a slightly weaker correlation compared to $f_{\phi u'_0}$. The positive correlation between $\phi$ and both the initial and final $u'$ corroborates that the cumulative control $\phi$ achieved through the optimization process considers minimizing wall friction across the entire time horizon. Conversely, in terminal wall friction control (with $\Delta t^+=50$, figure \ref{fig:oppo_phiu}c), the joint PDF $f_{\phi u'_0}$ (colored contour) is symmetric around $u'=0$, indicating that $\phi$ is largely uncorrelated with the initial $u'_0$. This observation is consistent with figure \ref{fig:phi_cf}(b). Regarding $f_{\phi u'_{\Delta t}}$ (dashed isolines), a distinct positive correlation is observed, confirming that terminal control of wall friction exclusively focuses on reducing friction at the terminal time.

The joint PDFs $f_{\phi u'_0}$ and $f_{\phi u'_{\Delta t}}$ of cumulative TKE control depicted in figure \ref{fig:oppo_phiu}(b) demonstrates a strong positive correlation between the control variable $\phi$ and both $u'_0$ and $u'_{\Delta t}$. In contrasted with cumulative wall friction control, these distributions are more tightly centered around the line given by $\phi/\sigma_\phi=u'/\sigma_{u'}$, indicating a direct reliance of $\phi$ on $u'$. Similarly, for terminal TKE control, control $\phi$ also displays a strong positive correlation with $u'_{\Delta t}$. While $\phi$ shows a weaker correlation with $u'_0$, there is still a distinct skew of $f_{\phi u'_0}$ towards the first and third quadrants. These observations align with the instantaneous field data shown in figure \ref{fig:k_phi}(b,d). 

It is worth reiterating the distinction between terminal wall friction control and terminal TKE control. The control variable $\phi$ for the former has almost no correlation with the initial $u'_0$ condition, whereas the control $\phi$ for latter shows a notable correlation with $u'_0$. Although both scenarios only incorporate the terminal state in the loss function, this discrepancy arises from the inherent characteristics of the target, including spatial extent, wall location, and related variables. TKE is an appealing option as it pertains to every component of turbulent velocity throughout the channel. Nonetheless, several issues remain unresolved, such as why friction control optimization underperforms compared to TKE control, and what fundamentally differentiates the control strategies of wall friction and TKE control. These concerns will be addressed through a more detailed examination of the statistics of the turbulent channel.

\begin{figure}
    \centering
    \includegraphics[width=1\textwidth]{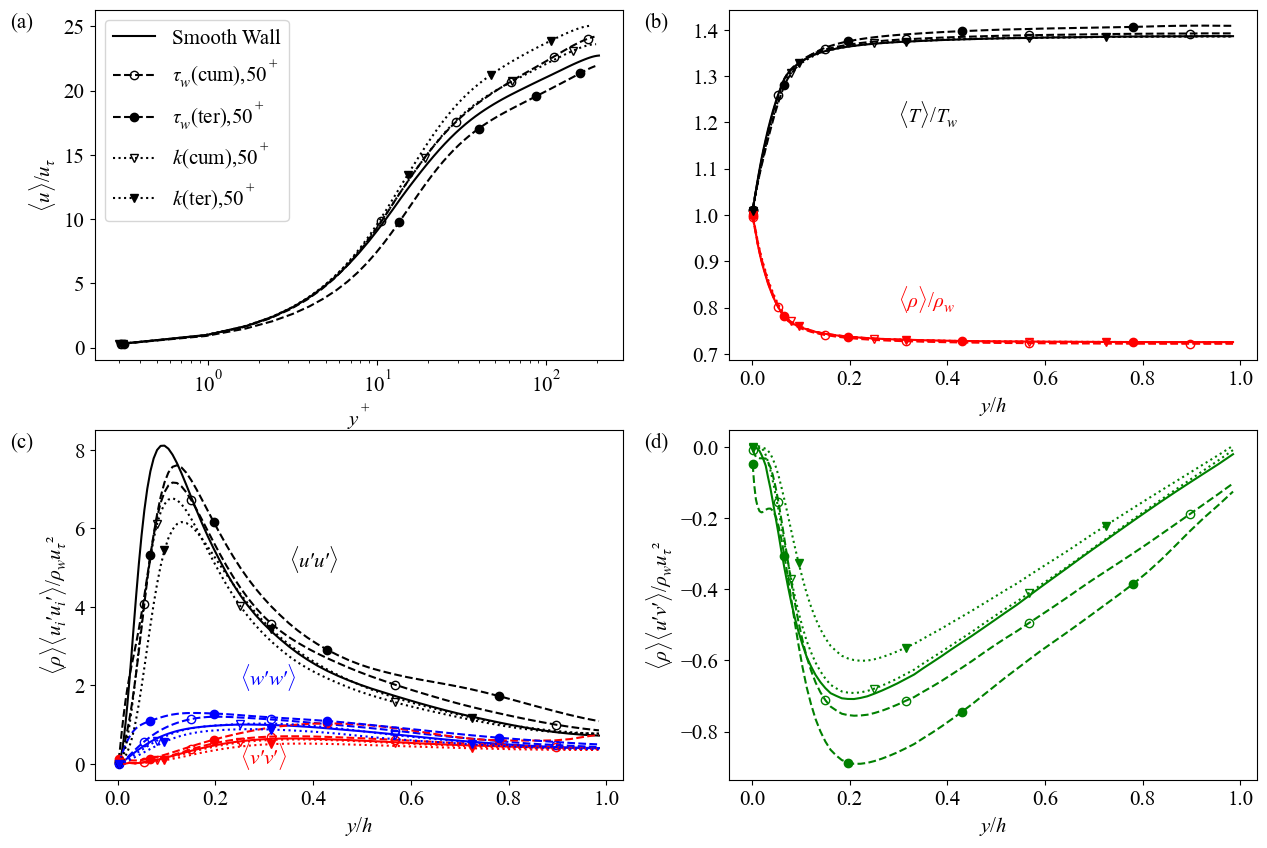}
    \caption{Mean statistics of opposition control cases compared with smooth wall channel. (a) mean streamwise velocity $\langle u\rangle$;(b) mean temperature $\langle T\rangle$ and mean density $\langle \rho\rangle$; (c) turbulent intensity $\langle u_i'u_i'\rangle$; (d) Reynolds stress $\langle u'v'\rangle$.}
    \label{fig:stat_oppo}
\end{figure}
\subsubsection{The mean statistics}
This section delves into the mean statistics of controlled flow fields to understand the control mechanisms aimed at different targets. Table \ref{tab:oppo_cf} provides a summary of the wall friction coefficient $C_f$ for a standard turbulent channel, along with the relative changes $\Delta C_f$ observed in the control scenarios. In each case, the skin friction is calculated using the flow fields over the time interval $t^+=500-2000$. Currently, our focus is on the overall wall friction coefficient, while the decomposition of wall friction will be addressed later in the section. With the exception of the terminal $\tau_w$ control cases, which exhibit a substantial rise in wall friction owing to instability, all opposition control methods result in varying levels of drag reduction. The differences in effectiveness across different control targets can be somewhat elucidated by examining the mean statistics.

Figure \ref{fig:stat_oppo}(a) displays the average statistics for the four typical scenarios with $\Delta t^+=50$, including both cumulative and terminal control of wall friction and TKE. The mean streamwise velocity profiles $\langle u\rangle/u_\tau$ (figure \ref{fig:stat_oppo}a) show an upward shift in the outer region for all controlled cases when compared to the reference smooth wall scenario. This upward shift signifies a decrease in wall friction velocity $u_\tau$, which is directly associated with wall shear stress $\tau_w$. This can be confirmed by comparing to table \ref{tab:oppo_cf}.  The extent of drag reduction tends to correspond with the degree of shift in the $\langle u\rangle/u_\tau$ profile. From the extent of the shift, the terminal TKE control demonstrates superior performance in friction reduction among the four cases. 

The average density $\langle \rho\rangle$ profiles (figure \ref{fig:stat_oppo}a) in all cases remain largely unchanged, indicating that the compressibility of the channel is not affected by opposition control. Similarly, the temperature distribution $\langle T\rangle$ within the channel exhibits minimal alteration. Only the case involving friction control of the terminal wall shows a slight increase in temperature in the central area of the channel, which could be related to the increase in viscous dissipation owing to a notable increase in the intensity of the turbulence (Figure \ref{fig:stat_oppo}c).

In the scenarios involving wall friction control, the peak of $\langle u'u' \rangle$ is seen to move farther from the wall. Although the peak closer to the wall decreases, there is an increase in $\langle u'u' \rangle$ within the outer region. Moreover, a notable rise in $\langle v'v' \rangle$ and $\langle w'w' \rangle$ is detected in the terminal friction control case. In contrast, for the TKE control cases, $\langle u'u' \rangle$ is not only significantly reduced near the wall but also maintains a low level in the outer region. Similarly, $\langle v'v' \rangle$ and $\langle w'w' \rangle$ are diminished in TKE control cases.

Similarly to the intensity of the turbulence, the Reynolds stress $\langle u'v' \rangle$ (figure \ref{fig:stat_oppo}d) of $\tau_w$ control cases shows negligible reductions near the wall, yet it is considerably greater than in the smooth wall scenario in the outer region. In the TKE control cases,  particularly with terminal control, the $\langle u'v' \rangle$ magnitude are notably lower than the baseline level.

Analysis of the average statistics indicates a fundamental contrast in the approaches of wall friction control versus TKE targeted control. In cases aimed at controlling wall friction, energetic turbulent structures such as streaks and vortices are displaced from the wall, resulting in an outward shift of the turbulence intensity and Reynolds stress peaks. Although this method may be temporarily effective, it induces additional turbulence fluctuations in the outer regions, ultimately enhancing the mixing and undermining the friction reduction efforts. Conversely, TKE control not only relocates the turbulence but also suppresses it without creating additional disturbances. The different tactics employed by the two control targets are related to the scope of their information. Wall friction control focuses solely on wall-specific data and lacks insight into outer region dynamics. Consequently, its optimal strategy is to push the friction-inducing structures away. TKE control retains well rounded information across the entire channel, allowing it to successfully mitigate turbulence while avoiding any adverse effects.

\begin{table}
    \centering
\begin{tabularx}{1\textwidth}{>{\centering\arraybackslash}X >{\centering\arraybackslash}X >{\centering\arraybackslash}X >{\centering\arraybackslash}X >{\centering\arraybackslash}X }
        Reference Case &$C_{f,0}$ & $C_{f,0}^M$ & $C_{f,0}^F$ & $C_{f,0}^T$ \\
       Smooth Wall & 7.84$\times10^{-3}$ & 4.84$\times10^{-3}$ & 2.18$\times10^{-5}$ & 2.99$\times10^{-3}$ \\\\ %\hline
       Control Cases &$\Delta C_f/C_{f,0}$ & $\Delta C_f^M/ C_{f,0}^M$ & $\Delta C_f^F/ C_{f,0}^F$ & $\Delta C_f^T/ C_{f,0}^T$\\
       $\tau_w$(cum),$25^+$ & -12.5\% & -11.0\% & -34.8\% & -15.6\% \\
       $\tau_w$(cum),$50^+$ & -9.5\% & -9.1\% & -33.2\% & -11.1\% \\
       $\tau_w$(ter),$25^+$ & +13.8\% & -14.2\% & -20.8\% & +55.8\% \\
       $\tau_w$(ter),$50^+$ & +19.5\% & -20.0\% & -19.3\% & +80.0\% \\
        $k$(cum),$25^+$ & -8.7\% & -4.8\% & -28.6\% & -18.6\% \\
       $k$(cum),$50^+$ & -8.3\% & -3.0\% & -31.4\% & -17.5\% \\
       $k$(ter),$25^+$ & -11.8\% & -4.7\% & -38.8\% & -26.8\% \\
       $k$(ter),$50^+$ & -18.2\% & -10.7\% & -52.2\% & -33.4\% \\ 
\end{tabularx}
\caption{The wall friction of the turbulent channel and the relative change in opposition control cases. }
    \label{tab:oppo_cf}
\end{table}

\subsubsection{Decomposition of wall friction}
To further understand the change of friction source in different control cases, we apply skin-friction drag decomposition for compressible channel flow proposed by \cite{Renard_Deck_2016} and \cite{li2019decomposition},
\begin{equation}
\begin{split}
        C_f=\underbrace{\frac{2}{\rho_b u_b^3} \int_0^h\langle\mu\rangle \frac{\partial\langle u\rangle}{\partial y} \frac{\partial\{u\}}{\partial y} \mathrm{~d} y}_{C_{f}^M}+\underbrace{\frac{2}{\rho_b u_b^3} \int_0^h\left\langle\mu^{\prime} \frac{\partial u^{\prime}}{\partial y}+\mu^{\prime} \frac{\partial v^{\prime}}{\partial x}\right\rangle \frac{\partial\{u\}}{\partial y} \mathrm{~d} y}_{C_{f}^F}\\
        +\underbrace{\frac{2}{\rho_b u_b^3} \int_0^h\langle\rho\rangle\left\{-u^{\prime \prime} v^{\prime \prime}\right\} \frac{\partial\{u\}}{\partial y} \mathrm{~d} y}_{C_{f}^T}.
\end{split}
\label{eq:RD}
\end{equation}
where $C_f^M$ and $C_f^F$ are both related to the direct molecular viscous dissipation, which transforms the power of the skin friction drag into heat. $C_f^M$ dependent on the mean shearing in the flow and $C_f^F$ is generated due to the thermodynamic fluctuations in compressible flow. $C_f^T$ represents the power converted into the production of turbulent kinetic energy induced by turbulent fluctuations.

Although this decomposition was originally formulated based on the no-slip wall condition, it is also applicable to opposition control with zero-net-flux at the wall. For all the cases, the residual error of RD identity is below $\pm2.3\%$. The RD identities for all cases are documented in Table \ref{tab:oppo_cf}. Across all scenarios, the primary contributors are $C_f^M$ and $C_f^T$, while the contribution from the compressibility effect of the flow, $C_f^F$, is minimal due to the low Mach number. Figure 10 presents a comparison between the pre-multiplied integrands of the four typical cases (as in figure \ref{fig:oppo_phiu} and \ref{fig:stat_oppo}) and a smooth-wall channel. 

\begin{figure}
    \centering
    \includegraphics[width=1\linewidth]{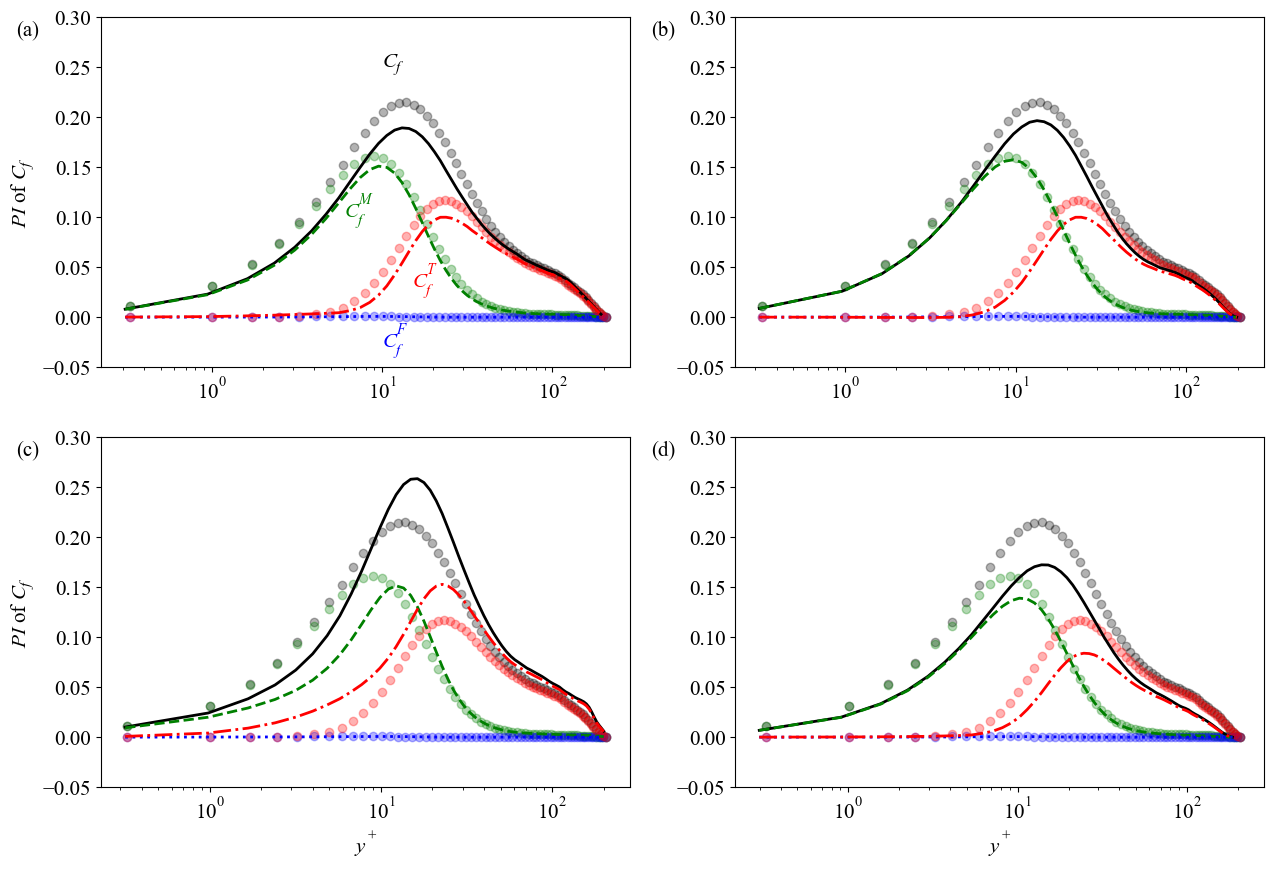}
    \caption{Pre-mulitplied integrands (PI) of RD identity as a function of $y^+$ in opposition control cases. $C_f^M$, $C_f^F$ and $C_f^T$ are denoted by dashed lines, dotted lines and dash-dotted lines, respectively. The PIs of smooth wall channel are superimposed with circles. (a) cumulative $\tau_w$ control, $\Delta t^+=50$; (b) cumulative $k$ control, $\Delta t^+=50$; (c) terminal $\tau_w$ control, $\Delta t^+=50$; (d) terminal $k$ control, $\Delta t^+=50$.}
    \label{fig:RD_oppo}
\end{figure}

In the case of cumulative $\tau_w$ control with $\Delta t^+=50$ (figure \ref{fig:RD_oppo}a), there is an approximate 10\% decrease in both $C_f^M$ and $C_f^T$. The decrease in $C_f^T$ is related to the positive correlation between the control parameter $\phi$ and the fluctuating velocity $u'$ as illustrated in figure \ref{fig:oppo_phiu}(a).  The decrease in $C_f^M$ suggests alterations in the mean shear close to the wall, particularly for $y^+$ below 10. This change is so subtle that it is not prominently observable in instantaneous flow fields (figure \ref{fig:oppo_flow}a,c,e) or the normalized mean velocity profile (figure \ref{fig:stat_oppo}a).
The terminal $\tau_w$ control (figure \ref{fig:RD_oppo}c) more intensely altered the mean flow near the wall, decreasing $C_f^M$ by 20\%. However, the additional fluctuations it induced considerably amplified turbulent production, increasing $C_f^T$ by 80\%. Consequently, its impact on mean shear at the wall is completely cancelled by increased turbulence, leading to increased wall friction.

In the cumulative $k$ control illustrated in Figure \ref{fig:stat_oppo}(b), the integrand of $C_f^T$ is primarily reduced in the near-wall region ($y^+<30$), while $C_f^M$ remains mostly unaffected. Conversely, the terminal $k$ control takes a notably more aggressive approach, significantly diminishing turbulence production near the wall and in the outer region, resulting in a 33.4\% decrease in $C_f^T$. Additionally, the mean shear on the wall undergoes a considerable decline. The effective reduction in both $C_f^T$ and $C_f^M$ positions this as the most efficient opposition control strategy achieved through the current optimization.

The findings from RD identity support our previous hypothesis on the distinct mechanisms employed in $\tau_w$ targeting control as opposed to $k$ targeting control. In cases of $\tau_w$ control, which only use limited data at the wall, the strategy tends to reduce shear at the wall, albeit at the cost of introducing flow disturbances. In contrast, $k$ control cases, equipped with a comprehensive dynamic flow profile within the loss function, adopt a turbulence cancellation strategy. The latter approach is evidently more efficient. Although auto-differentiation offers a rapid method for obtaining gradients with a vast number of parameters, the final outcomes are still heavily influenced by the selection of the optimization target and loss function, which must be grounded in physical understanding.

\subsection{The performance of tunable permeable wall}

\begin{figure}
    \centering
    \includegraphics[width=1\linewidth]{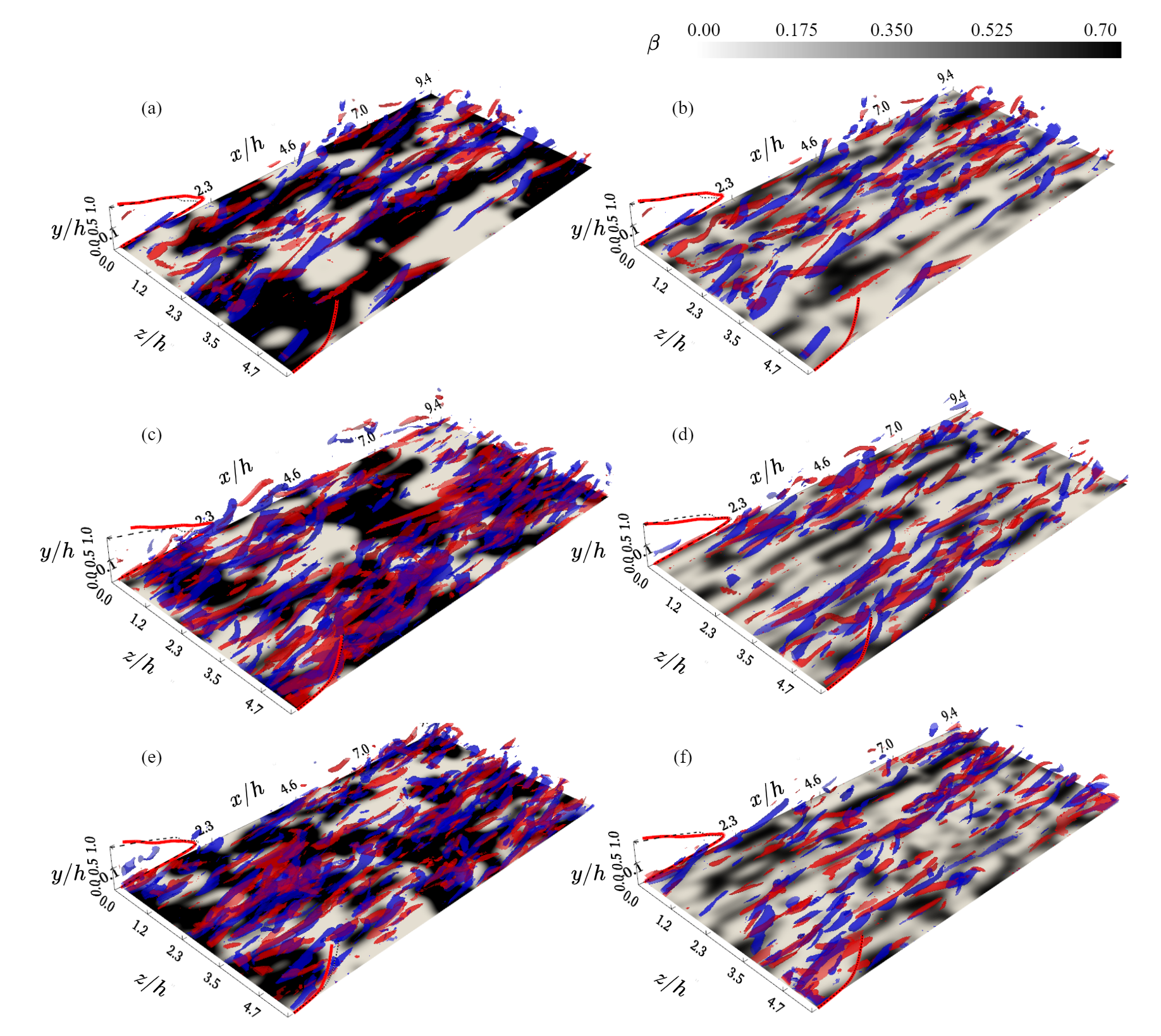}
    \caption{Vorticity structures above tunable permeable wall. Panels (a,c,e) show cumulative wall friction control with $\Delta t^+=25$, while panels (b,d,f) depict terminal TKE control with the same $\Delta t^+$. The pairs of panels (a,b), (c,d), and (e,f) represent $t^+=50$, $t^+=500$, and $t^+=2000$ from the onset of control, respectively. Blue and red iso-surfaces indicate streamwise vorticity at $\omega_x=\pm\sigma_{\omega_x}$, $\sigma_{\omega_x}$ being the standard deviation of $\omega_x$ at $t=0$. Control $\phi$ is illustrated on the wall with colored contours. Present snapshot profiles of $\langle u\rangle$ and $-\langle u'v'\rangle$ are overlaid on front ($z=1.5\pi$) and back ($z=0$) planes with solid red lines. Black dashed lines show profiles from the smooth wall scenario for comparison. See also supplementary videos 3 and 4 for the full simulation duration.}
    \label{fig:porous_flow}
\end{figure}

The main drawback of opposition control, similar to other active control techniques, is its high energy consumption. Consequently, passive control strategies continue to be widely intriguing for industrial implementation. In this research, we investigate the effectiveness of a permeable wall for drag reduction. Prior research has indicated that a stationary permeable wall can reduce drag if the permeability is extremely low and predominantly aligned in the streamwise direction \cite[]{Rosti.2018}. Aside from this specific instance, most studies on permeable walls have noted an increase in drag \cite[]{Abderrahaman.2017,wang2021information, wang2021anassess}. We are now removing the temporal invariance constraint on the permeable wall to allow the permeability to vary with each optimization cycle. Although walls with tunable permeability are not frequently explored, it can be implemented using mechanical systems, shape-memory alloys, or advanced materials. 

Similarly to opposition control, in each optimization episode, we adjust the boundary condition parameter ($\beta$ for the permeable wall) according to our control objectives. However, unlike opposition control, the permeable wall does not dynamically add or remove mass from the flow. Instead, it solely modulates local flux through changes in local permeability, which may considerably limit its ability to alter the channel. However, it is intriguing to determine whether the high degrees of freedom offered by tunable permeability can be leveraged to achieve drag reduction.

Similar to opposition control, there are several key factors to take into account here, such as the target variable (either wall friction $\tau_w$ or TKE $k$), the type of loss function (whether cumulative or terminal), and the chosen time horizon ($\Delta t^+=25$ or $\Delta t^+=50$). To illustrate the evolution of turbulent structures over a permeable wall, figure \ref{fig:porous_flow} presents the instantaneous vorticity structures for two representative scenarios: cumulative $\tau_w$ control with $\Delta t^+=25$ and terminal $k$ control with $\Delta t^+=25$. The flow within the channel undergoes substantial alteration by intermittently varying the distribution of $\beta$ (represented by the gray contour on the plane $y=0$). For the cumulative $\tau_w$ control case (figure \ref{fig:porous_flow}a,c,e), the turbulence is intensified with obvious increase of the population of vortices. Correspondingly, there is an increase of Reynolds stress (red solid line in plane $z=0$). In the terminal $k$ control case (figure \ref{fig:porous_flow}b,d,f), the number of vortices in the fields is slightly reduced. This is also evidenced by the decrease of Reynolds stress (red solid line in plane $z=0$) in the outer region. 

The $\beta$ fields (depicted as gray contours on the $y=0$ plane) predominantly exhibit two distinct values. For most parts of the wall, the local $\beta$ registers either as 0, indicating non-permeability, or 0.7, which represents the upper bound for $\beta$ in the current optimization study. Regions exhibiting $\beta$ values in between these extremes are relatively small. In the scenario of cumulative $\tau_w$ control (see figure \ref{fig:porous_flow}a,c,e), the permeable zone forms an interconnected network covering a substantial portion of the wall. Conversely, for the terminal $k$ control case, the permeable zones appear more segregated and streak-like. A notable increase in intermediate $\beta$ values suggests the necessity for more refined controls to manage turbulence. These distinctions in the patterns of the $\beta$ map controls highlight fundamentally different control mechanisms between the $\tau_w$ and $k$ control scenarios. Further discussion on these differences will be provided in subsequent sections.

\subsubsection{The development of wall friction on tunable permeable walls}
\begin{figure}
    \centering
    \includegraphics[width=0.8\linewidth]{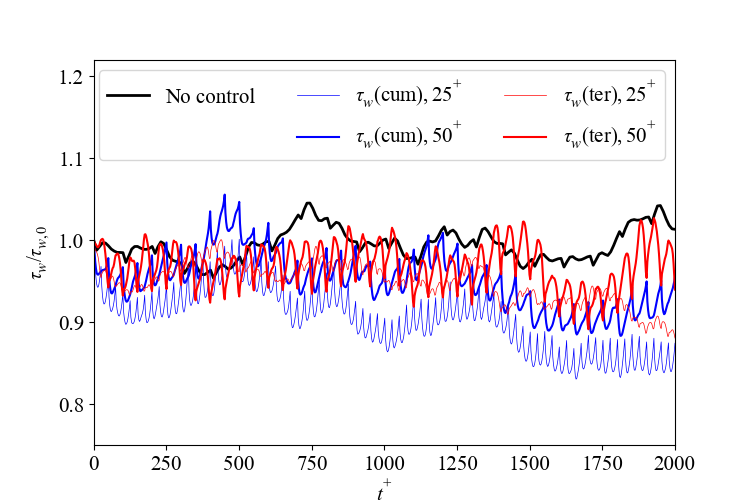}
    \caption{History of wall friction over tunable permeable wall targeting loss functions associated with $\tau_w$.}
    \label{fig:porous_drag}
\end{figure}
Figure \ref{fig:porous_drag} presents the progression of wall friction under the influence of wall friction control. With the permeable wall boundary condition, both cumulative and terminal $\tau_w$ control achieve stable outcomes in drag reduction, even in the terminal control scenarios. Notably, cumulative $\tau_w$ control with a time horizon $\Delta t^+=25$ achieves over 10\% drag reduction, while other control scenarios yield roughly a 5\% reduction. 

It is noteworthy that the permeable wall, despite its lack of active flow manipulation ability like opposition control, possesses an advantage in maintaining stable control outputs with terminal control, relying on minimal information from the near-wall position. This is partly due to the special properties of disturbances generated by the permeable wall. The wall-normal flux at the permeable boundary is restricted to be proportional to the pressure fluctuations, which are mostly chaotic, with an upper limit for $\beta$ set at 0.7. These conditions ensure that the wall-normal flux remains limited and prevents excessive interfacial flux. Moreover, the drag reduction mechanism of the permeable wall differs from that used in opposition control, a topic to be further discussed in a subsequent section. 

\begin{figure}
    \centering
    \includegraphics[width=0.8\linewidth]{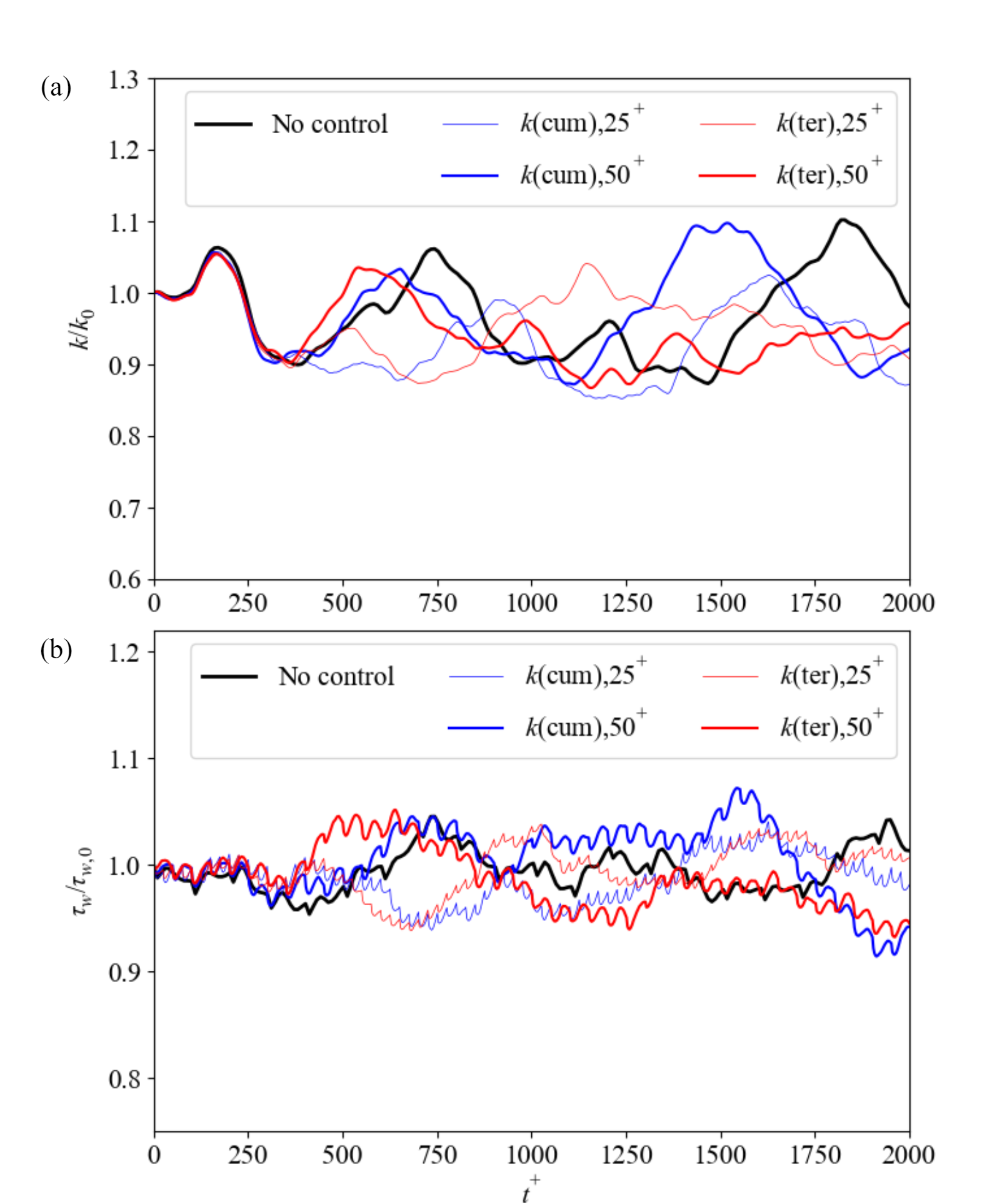}
    \caption{The history of (a) TKE $k$ and (b) wall friction $\tau_w$ with tunable permeable wall targeting TKE related loss function.}
    \label{fig:porous_TKE}
\end{figure}

In the case of TKE control, the use of permeable walls proves to be less efficient than opposition control. Figure \ref{fig:porous_TKE} illustrates the TKE and wall friction over time with permeable wall intervention. Generally, the reduction in TKE and wall friction is insignificant. The most successful scenario is terminal $k$ control with $\Delta t^+=50$, achieving around 5\% suppression in TKE and a 2\% reduction in drag. It is not surprising that the permeable wall is ineffective at controlling TKE. The current permeable wall boundary condition only includes a wall-normal component. Previous studies \cite[]{Rosti.2018} have indicated that even slight wall-normal permeability can considerably enhance turbulence production and wall friction. Nevertheless, we demonstrate that reducing TKE and drag is indeed achievable through an optimized and adjustable distribution of $\beta$.

\subsubsection{The characteristics of $\beta$ distribution}

\begin{figure}
    \centering
    \includegraphics[width=1\linewidth]{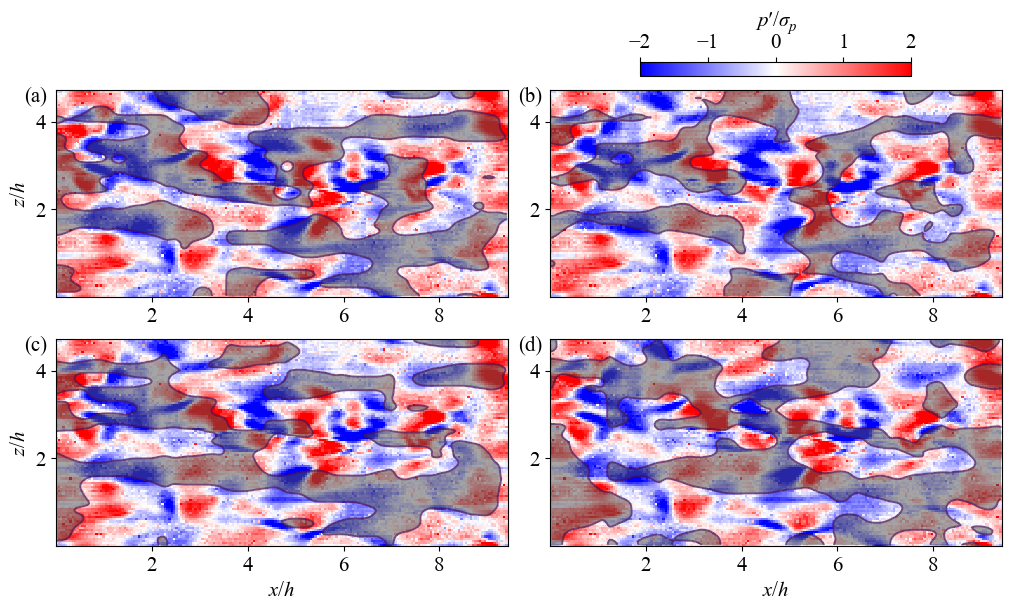}
    \caption{The $\beta$ map of the permeable wall for wall friction control with different targets and time horizons. (a) cumulative control with $\Delta t^+=25$; (b) terminal control with $\Delta t^+=25$; (c) cumulative control with $\Delta t^+=50$;(d) terminal control with $\Delta t^+=50$. The gray patches are permeable regions with $\beta\ge0.4$. The color map shows the $p'$ at buffer layer $y^+=15$.}
    \label{fig:porous_cf_beta}
\end{figure}

\begin{figure}
    \centering
    \includegraphics[width=1\linewidth]{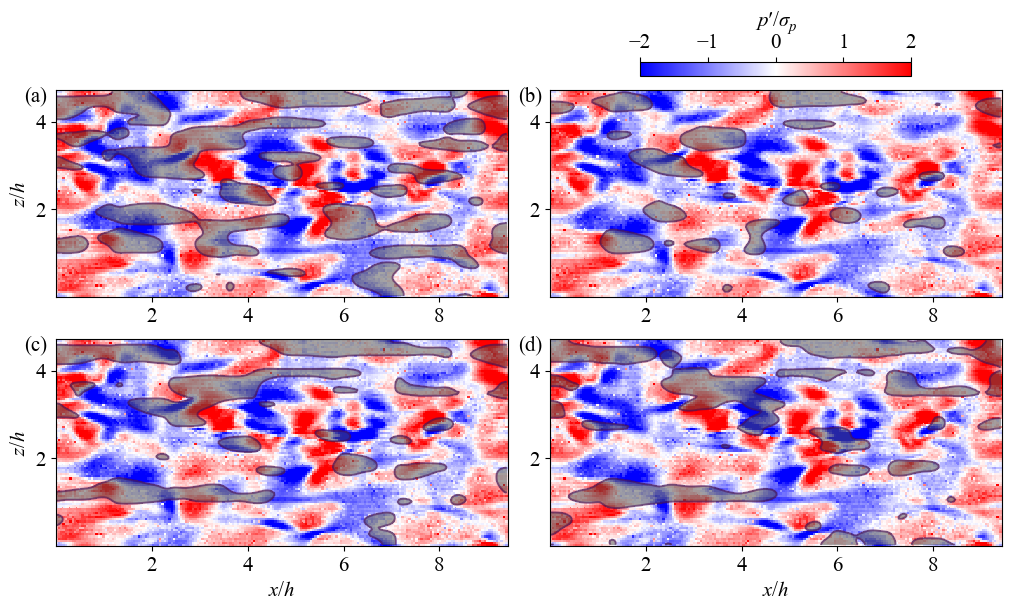}
    \caption{The $\beta$ map of the permeable wall for TKE control with different targets and time horizons. (a) cumulative control with $\Delta t^+=25$; (b) terminal control with $\Delta t^+=25$; (c) cumulative control with $\Delta t^+=50$;(d) terminal control with $\Delta t^+=50$. The other setting are the same as figure \ref{fig:porous_cf_beta}.}
    \label{fig:porous_k_beta}
\end{figure}

\begin{figure}
    \centering
    \includegraphics[width=1\linewidth]{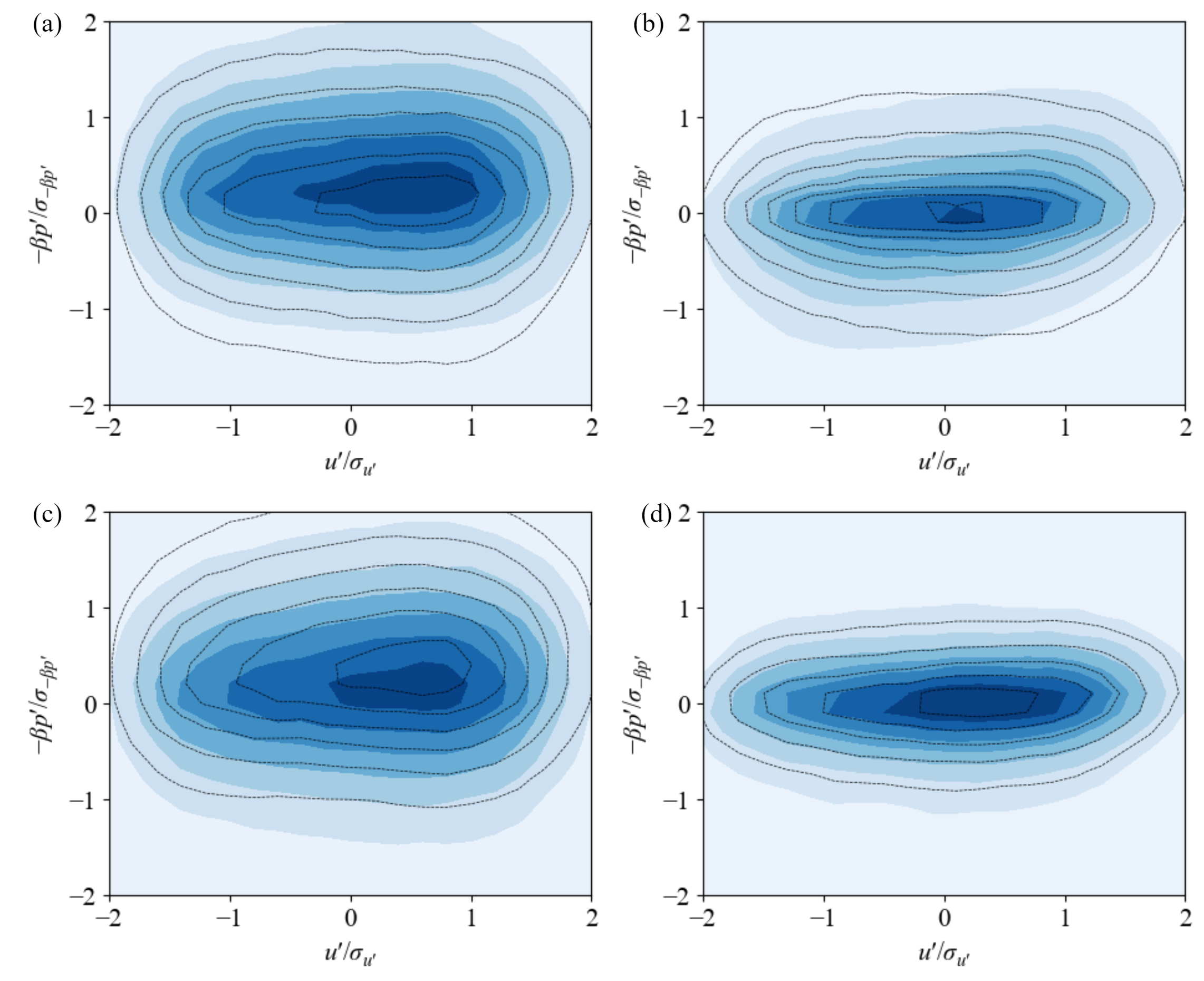}
    \caption{The joint probability density function (PDF) $f_{-\beta p', u'}$ between the permeable wall flux and streamwise fluctuation $u'$. (a) cumulative control of wall friction with $\Delta t^+=25$; (b) cumulative control of TKE with $\Delta t^+=25$;(c) terminal control of wall friction with $\Delta t^+=25$; (d) terminal control of wall TKE with $\Delta t^+=25$. The colored contour show the joint PDF $f_{-\beta p'_0, u'_0}$ between wall flux and streamwise fluctuation $u'$ at the initial time of the episodes, and the dashed isolines show the joint PDF $f_{-\beta p'_{\Delta t}, u'_{\Delta t}}$ at terminal time of the episodes.}
    \label{fig:porous_betau}
\end{figure}

The instantaneous flow fields depicted in figure \ref{fig:porous_flow} have demonstrated the distinctions between the $\beta$ fields in the $\tau_w$ and $k$ control scenarios. This section delves deeper into the characteristics of the $\beta$ field to gain a clearer understanding of how the permeable wall adjusts to the flow field, thereby achieving drag reduction and minimizing turbulence. 

Figure \ref{fig:porous_cf_beta} presents the $\beta$ fields for the $\tau_w$ control scenarios, using a gray color map. As noted in figure \ref{fig:porous_flow}, the $\beta$ fields predominantly exhibit two values, 0 and 0.7, which correspond to non-permeable and permeable zones. In figure \ref{fig:porous_flow}, permeable regions where $\beta\ge 0.4$ are depicted by gray patches. Additionally, the $p'$ field from the initial state of the current episode is overlaid, allowing the inference of local $v'$ at the wall.

For both cumulative and terminal control scenarios, the $\beta$ map reveals a pattern that is consistent across cases. In scenarios with an extended time horizon, the permeable zones exhibit greater interconnectivity, and the primary areas remain comparable. In all instances, there is noticeable overlap between the permeable zone and the negative $p'$ region. This indicates that these $\beta$ map distributions generate a net positive wall-normal flux. The injection of additional momentum into the near-wall region can modify the velocity gradient and the characteristics of turbulence, partially elucidating the mechanism behind drag reduction. This also clarifies why the $\beta$ map possesses two distinct values. Strategically positioning permeable zones, especially in regions with negative $p'$, to have maximum permeability supports maximal upward fluid movement, while setting non-permeable zones is crucial to entirely block any unwanted downward momentum.

Figure \ref{fig:porous_k_beta} illustrates the $\beta$ field in $k$ control cases. Comparing to the $\tau_w$ control cases in figure \ref{fig:porous_cf_beta}, the permeable areas in $k$ control cases are smaller and more scattered, suggesting that the TKE control has more stringent requirement on the wall normal disturbances on the wall. Moreover, no obvious correlation between $\beta$ and $p'$ can be observed, hence the there is no net momentum flux across the wall.

The statistical features of the $\beta$ field are further elucidated in figure \ref{fig:porous_betau}, which depicts the joint probability density function (PDF) $f_{-\beta p',u'}$ between the vertical velocity at the permeable wall, represented by $v_w=-\beta p'$, and the streamwise fluctuation $u'$ within the buffer layer ($y^+=15$). In contrast to figure \ref{fig:oppo_phiu}, color contours in this figure show the joint PDF at the start of the episodes, labeled $f_{-\beta p'_0,u'_0}$, while dashed isolines illustrate the joint PDF at the episodes' endpoint, labeled $f_{-\beta p'_{\Delta t},u'_{\Delta t}}$. With cumulative $\tau_w$ control (figure \ref{fig:porous_betau}a), the initial joint PDF $f_{-\beta p'_0,u'_0}$ prominently biases towards the positive $-\beta p'$ direction, suggesting a positive wall-normal flux. At the terminal time, a smaller bias is noted in $f_{-\beta p'_{\Delta t},u'_{\Delta t}}$. Conversely, under terminal $\tau_w$ control (figure \ref{fig:porous_betau}c), the terminal joint PDF $f_{-\beta p'_{\Delta t},u'_{\Delta t}}$ exhibits a more pronounced bias to the positive $-\beta p'$ than the initial joint PDF $f_{-\beta p'_0,u'_0}$, implying that most of the wall-normal flux occurs at episodes' conclusion. For both cumulative and terminal $\tau_w$ control scenarios, the $f_{-\beta p',u'}$ displays symmetry around its horizontal centerline, indicating no significant correlation between these variables.

In the cumulative $k$ control (figure \ref{fig:porous_betau}b) , the wall-normal flux $-\beta p'$ poses a weak positive correlation with $u_0'$  with the shape of $f_{-\beta p'_0,u'_0}$ skews towards the first and third quadrants. Such positive correlation is not seen in the terminal joint PDF $f_{-\beta p'_{\Delta t},u'_{\Delta t}}$. A weak positive correlation between $-\beta p'$ and $u'$ is also observed in the joint PDFs in the terminal $k$ control
case (figure \ref{fig:porous_betau}d).

The joint PDFs in figure \ref{fig:porous_betau} confirms our observation in the instantanous $\beta$ fields in figure \ref{fig:porous_cf_beta} and \ref{fig:porous_k_beta}. For wall fricition control, the permeable wall adopted a completely different strategy than the opposition control, that is, exploiting the tunable permeability to inject net upward momentum in the near wall region in order to alter the shear at the wall. The TKE control, however, shares the same machanism with opposition control. In this case, wall-normal velocity induced by the permeable region is postively correlated to the $u'$ near the wall, hence cancelling the local Reynolds stress, which eventually leads to drag reduction.

\begin{table}
    \centering
\begin{tabularx}{1\textwidth}{>{\centering\arraybackslash}X >{\centering\arraybackslash}X >{\centering\arraybackslash}X >{\centering\arraybackslash}X >{\centering\arraybackslash}X 
>{\centering\arraybackslash}X 
>{\centering\arraybackslash}X}
       Control Cases &$\Delta C_f/C_{f,0}$ & $\Delta C_f^M/ C_{f,0}^M$ & $\Delta C_f^F/ C_{f,0}^F$ & $\Delta C_f^T/ C_{f,0}^T$ & $C_{f1}^V/ C_{f,0}$& $C_{f2}^V/ C_{f,0}$\\
       $\tau_w$(cum),$25^+$ & -15.8\% & -3.2\% & +5.4\% & +34.7\%& +14.7\%&-39.4\% \\
       $\tau_w$(cum),$50^+$ & -7.1\% & -2.5\% & +1.8\% & +12.5\%& +5.2\%&-17.1\% \\
       $\tau_w$(ter),$25^+$ & -4.9\% & -0.8\% & +4.2\% & +12.9\%& +4.7\%& -12.7\%\\
       $\tau_w$(ter),$50^+$ & -4.1\% & -0.3\% & +4.1\% & +18.5\% & -3.3\%& -5.7\%\\
        $k$(cum),$25^+$ & -2.3\% & -0.5\% & -14.6\% & -9.6\%& -4.1\%&+5.1\%\\
       $k$(cum),$50^+$ & -1.8\% & -0.8\% & -11.2\% & -7.9\%& -4.2\%&+5.4\%\\
       $k$(ter),$25^+$ & -1.4\% & -1.1\% & -8.9\% & -7.6\%& -4.5\%&+5.7\%\\
       $k$(ter),$50^+$ & -1.3\% & -0.9\% & -15.0\% & -8.0\% & -4.5\%&+5.7\%\\ 
\end{tabularx}
\caption{The the relative wall friction change in permeable wall cases. }
    \label{tab:porous_Cf}
\end{table}

\subsubsection{The mean statistics and wall friction decomposition}

To elaborate on how drag is reduced using a permeable wall, we analyze the flow field statistics in this section. Table \ref{tab:porous_Cf} provides a comparison of wall friction between the permeable wall channel and the smooth wall channel (refer to table \ref{tab:oppo_cf}). It is evident that only the cumulative $\tau_w$ control with $\Delta t^+=25$ achieves a 15\% reduction in drag, while the effectiveness of other methods remains under 8\%.

\begin{figure}
    \centering
    \includegraphics[width=1\linewidth]{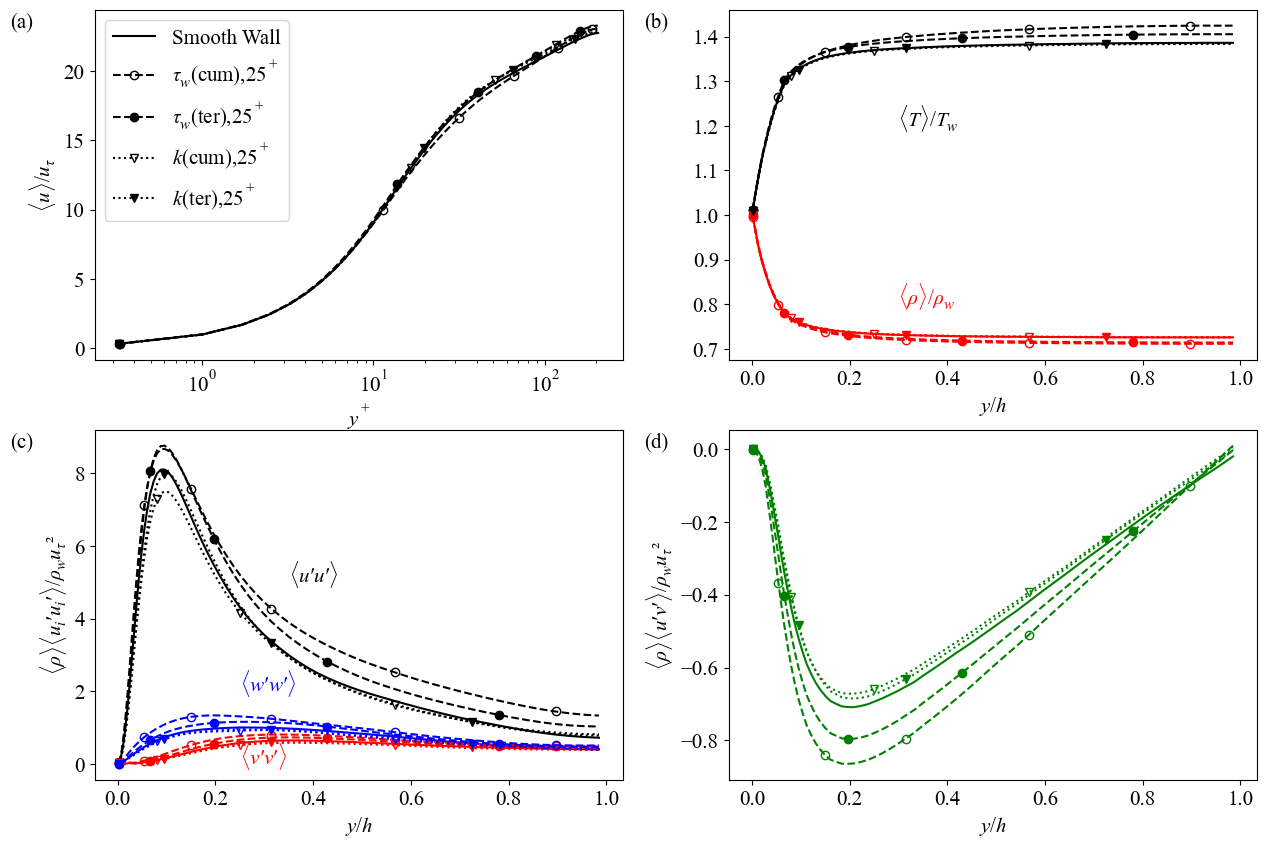}
    \caption{Mean statistics of permeable wall cases compared with smooth wall channel. (a) mean streamwise velocity $\langle u\rangle$;(b) mean temperature $\langle T\rangle$ and mean density $\langle \rho\rangle$; (c) turbulent intensity $\langle u_i'u_i'\rangle$; (d) Reynolds stress $\langle u'v'\rangle$.}
    \label{fig:stat_porous}
\end{figure}

Figure \ref{fig:stat_porous} presents the average statistics for four selected permeable wall cases. We chose the cases with shorter time horizon $\Delta t^+=25$, as they generally outperform the long time horizon ones. The mean streamwise velocity profiles (figure \ref{fig:stat_porous}a) remain largely unchanged due to the low drag reduction rate. In the $\tau_w$ control scenarios, the temperature at the channel's center (figure \ref{fig:stat_porous}b) shows a slight increase. This is attributed to a rise in turbulence intensity and Reynolds stress (figure \ref{fig:stat_porous}c,d), leading to more heat being dissipated into the channel and mildly reducing $\rho$. Conversely, in the $k$ control scenarios, both turbulence intensity and Reynolds stress exhibit a slight decrease.

To further understand the source of drag on permeable wall cases, we inspect the decomposition of wall friction. Note that for the permeable wall condition, the zero-net-flux condition does not hold anymore, therefore we have to re-derived the RD identity to include the additional terms generated by this. Here we show the RD identity for compressible channel with permeable wall. 
\begin{equation}
\begin{split}
       C_f=\underbrace{\frac{2}{\rho_b u_b^3} \int_0^h\langle\mu\rangle \frac{\partial\langle u\rangle}{\partial y} \frac{\partial\{u\}}{\partial y} \mathrm{~d} y}_{C_{f}^M}+\underbrace{\frac{2}{\rho_b u_b^3} \int_0^h\left\langle\mu^{\prime} \frac{\partial u^{\prime}}{\partial y}+\mu^{\prime} \frac{\partial v^{\prime}}{\partial x}\right\rangle \frac{\partial\{u\}}{\partial y} \mathrm{~d} y}_{C_{f}^F}\\
       +\underbrace{\frac{2}{\rho_b u_b^3} \int_0^h\langle\rho\rangle\left\{-u^{\prime \prime} v^{\prime \prime}\right\} \frac{\partial\{u\}}{\partial y} \mathrm{~d} y}_{C_{f }^T}\\
              +\underbrace{\frac{2}{\rho_b u_b^3} \int_0^h-\langle\rho\rangle(\{u\}-u_b)\{v\}\frac{\partial\{u\}  }{\partial y} \mathrm{~d} y}_{C_{f1}^V}\underbrace{-\frac{2\rho_w \{ v\}_w}{\rho_b u_b}}_{C_{f2}^V}.
\end{split}
\label{eq:RD_p}
\end{equation}
Comparing to the original RD identity equation (\ref{eq:RD}), two additional terms $C_{f1}^V$ and $C_{f2}^V$ emerged, representing the effect of mean convection in wall-normal direction. The first term $C_{f1}^V$ is the gain of mean streamwise kinetic energy in the absolute frame. The term $C_{f2}^V$ represents the effect of fluid injection at the permeable wall, showing that an positive $\{v\}$ flux at the wall would directly reduce the mean friction coefficient. The full derivation of equation (\ref{eq:RD_p}) is in appendix \ref{app:RD}. For the permeable wall cases, the residual error for the extended RD identity is confined within $\pm3.5\%$, which is slightly larger compared to the opposition control cases with the initial RD identity. This could be attributed to the limited sample time and the continuous momentum injection at the wall, which can violate the assumption of temporal homogeneity (see \ref{eq:RD_p}). However, the decomposition is still fairly reliable and may provide insight into wall friction over permeable walls.

Table \ref{tab:porous_Cf} lists the relative change in the components of RD identity for permeable wall cases. 
Note that for $C_{f1}^V$ and $C_{f2}^V$, the relative change is compare with the total friction. In the $\tau_w$ control scenarios, the main contributor to the reduction of drag is the wall flux term $C_{f2}^V$, which aligns with previous observations of a net positive $-\beta p'$ in figure \ref{fig:porous_betau}(a,c). However, the drag reduction attributed to wall flux is significantly counteracted by the increase in the kinetic energy of the turbulence $C_f^T$ and the kinetic energy of mean convection $C_{f1}^V$. Despite this, the cumulative $\tau_w$ control with $t^+=25$, with a maximum induced upward flux through the permeable wall, produces the highest drag reduction rate up to 15.8\%.

In all the TKE control cases, a slightly positive $C_{f2}^V$ is observed, which suggests a weak downward net flux at the wall. This minor magnitude is not clearly depicted in figure \ref{fig:porous_betau}(b,d). The presence of downward flux at the wall decreases the kinetic energy of both mean convection and turbulence, which results in a reduction in $C_{f1}^V$ and $C_{f}^T$. Additionally, the mean shear in the near wall region is altered, leading to a diminished $C_f^M$. Despite all these favorable changes to the flow, the overall drag reduction remain relatively minor compared to the $\tau_w$ control cases.

\begin{figure}
    \centering
    \includegraphics[width=1\linewidth]{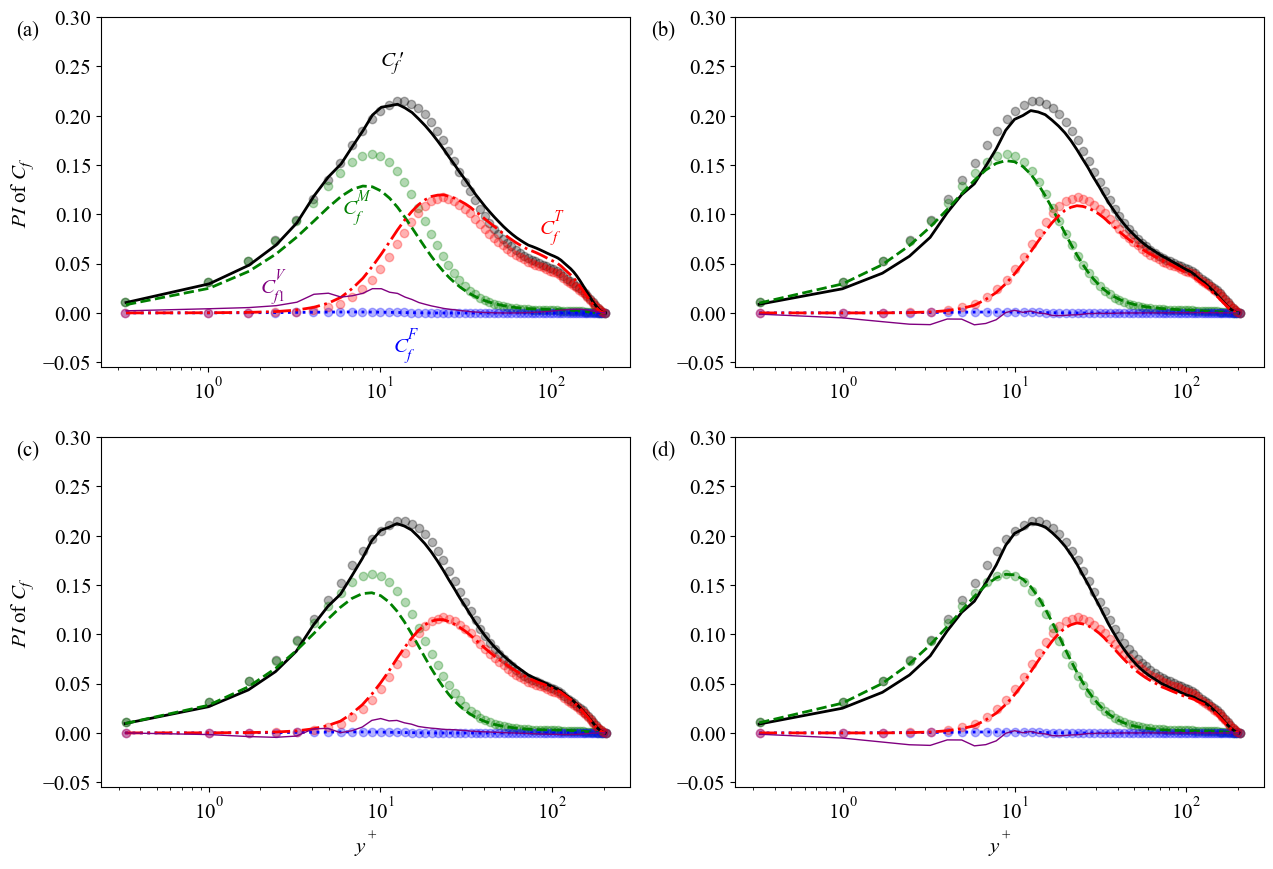}
    \caption{Pre-mulitplied integrands (PI) of RD identity as a function of $y^+$ for tunable permeable wall cases. $C_f^M$, $C_f^F$ , $C_f^T$ and $C_{f1}^V$ are denoted by dashed lines, dotted lines, dash-dotted lines and thin solid lines, respectively. The sum of all the integrands $C_f'$ are denoted by thick solid lines, where the contribution of $C_{f2}^V$ is not included. The PI of smooth wall channel is superimposed with circles of corresponding colors. (a) cumulative $\tau_w$ control, $\Delta t^+=25$; (b) cumulative $k$ control, $\Delta t^+=25$;(c) terminal $\tau_w$ control, $\Delta t^+=25$;(d) terminal $k$ control, $\Delta t^+=25$.}
    \label{fig:RD_porous}
\end{figure}

The premultiplied integrands of RD identity for the four typical cases in figure \ref{fig:stat_porous} is shown in figure \ref{fig:RD_porous}. Note that $C_{f2}^V$ is not in the form of integration and is not shown in figure \ref{fig:RD_porous}. For the $\tau_w$ control cases (figure \ref{fig:RD_porous}a,c), the mean shear term close to the wall is reduced. Meanwhile, the turbulent production term $C_f^T$ is enhanced. The net wall flux induces the vertical mean convection velocity, introduces extra $C_{f1}^V$ in the near wall area. This observation is consistent with our previous findings that the permeable wall in the $\tau_w$ control cases induces upward flux at the wall, altering the near wall shear stress and mean convect in the channel.   

As for the $k$ control cases (figure \ref{fig:RD_porous}b,d), both $C_f^T$ is reduced slightly, indicating the flux induced by permeable present an 'opposition control' effect on the turubence field. The $C_f^M$ is also slightly reduced. The downward wall flux also produces favorable $C_{f1}^V$.  However, these reduction sources are countered by $C_{f2}^V$, resulting in a marginal drag reduction result.

It appears that the optimization process effectively identified a `shortcut' for the $\tau_w$ control cases, which is indirectly injecting momentum into the boundary layer by adjusting the distribution of permeability. Although this method works well in an ideal scenario where the near-wall flow field is fully known, further refinement may be necessary for practical applications. Despite the fact that the drag reduction performance of tunable permeable wall is far below opposition control, we demonstrate that a permeable wall with only vertical permeability component can also achieve drag reduction when its local permeability is adjustable over time. It is also important to note that the adjustable permeable wall used in this study is not entirely a passive control method. Modulation of the wall permeability will undoubtedly consume energy. Bearing this in mind, cases with shorter episodes benefit from a higher energy input. This partly clarifies why the $t^+=25$ cases are generally more effective.

\section{Conclusive Remarks}\label{sec:conclusion}

In this study, we have developed an automatic differentiation (AD)-based optimization framework within a fully-differentiable solver for flow control in compressible turbulent channel flows. By developing a fully differentiable boundary condition, the exact gradients with respect to boundary control parameters are computed, enabling efficient optimization of flow control strategies. 
%This work represents a significant step forward in integrating differentiable fluid dynamics with gradient-based control, offering a streamlined and computationally efficient path for optimizing complex flow behaviors.

The study investigated two primary flow control strategies: opposition control and tunable permeable walls. We adopted the receding-horizon predictive control process, which consists of a series of `episodes'. In each optimization espisode, there were around $192\times96\times2\approx4\times10^4$ control variables and $192\times96\times128\times1200\approx3\times10^{9}$ state variables.
We explored the two control strategies under different optimization targets, specifically wall friction $\tau_w$ and turbulent kinetic energy (TKE), across varying time horizons. Opposition control strategies targeting TKE consistently outperform those directly minimizing $\tau_w$ in terms of stability and effectiveness. Specifically, terminal TKE control with $\Delta t^+=50$ manages to reduce drag by approximately 20\% by diminishing turbulence intensity across the entire channel domain.

In contrast, opposition control directly targeting $\tau_w$ appears to be less effective. Although this strategy initially showed a promising result in the initial phase of the control, it suffered from instability over longer time scales, particularly with terminal $\tau_w$ controls. This instability comes from the tendency of the $\tau_w$ control to push energetic turbulence structures away from the wall, leading to an increase in turbulent fluctuations in the outer flow region. The resulting redistribution of turbulence often led to an overall increase in Reynolds stresses, undermining the effectiveness of friction reduction. These findings highlight the importance of selecting loss functions that include information from the entire flow field to ensure stable and effective control outcomes.

We also explored the potential of using a tunable permeable wall as a quasi-passive control mechanism. Unlike active methods such as opposition control, the tunable permeable wall does not directly inject or remove mass directly from the flow. Instead, it modulates the flow dynamics through spatial variations in wall permeability. This performance of this strategy appears to be quite stable, even with terminal $\tau_w$ control cases. This stability is attributed to its naturally constrained wall-normal flux, which prevents excessive disturbances and ensures a gradual and mild control input.

The permeable wall demonstrated the ability to achieve up to 15\% drag reduction using cumulative $ \tau_w$ as loss functions. This is achieved by consistently inducing upward flux by adjusting the local permeability, which adds momentum into the near-wall region. It is proven with wall friction decomposition that this upward momentum contributes directly to drag reduction. However, the effectiveness of the permeable wall in reducing TKE was notably limited compared to the opposition control. This is likely due to its inherent limitations in dynamically adjusting the flow field, as the boundary condition only allows controlled leakage rather than active momentum injection. Nontheless, the stable reduction of drag using an adjustable permeable wall has been seldom investigated, and it presents a novel opportunity for the application of porous materials in industrial settings.

%The integration of AD into fluid dynamics simulations offers a powerful tool for optimizing flow control, particularly in compressible turbulent flows where conventional optimization methods may struggle due to the high dimensionality and nonlinearity of the problem. By automating the gradient computation process, the AD-based approach significantly reduces the computational cost and complexity associated with traditional adjoint methods, making it more accessible for a wider range of flow control applications.

The integration of AD into fluid dynamics simulations offers a powerful tool for optimizing flow control with high dimensionality and nonlinearity. The AD-based approach significantly reduces the complexity associated with traditional adjoint methods, making it more accessible for a wider range of flow control and optimization applications. The findings of this study suggest several avenues for future research. First, extending the optimization framework to higher Mach number flows could provide valuable insights into the control of supersonic and hypersonic boundary layers, where thermal and aerodynamic effects are more pronounced. Moreover, the potential to integrate AD-based optimization with machine learning techniques, such as reinforcement learning, could enable adaptive control strategies that dynamically adjust to changing flow conditions in real time.

\begin{acknowledgements}
%The study is funded by Deutsche Forschungsgemeinschaft (DFG, German Research Foundation) project SFB-1313 (project No. 327154368) and under Germany’s Excellence Strategy-EXC2075-390740016. 
%We acknowledge the support by the Stuttgart Center for Simulation Science (SimTech). 
The authors report no conflict of interest. This work is supported by the Fundamental Research Funds for the 
Central Universities (China). All authors gratefully acknowledge the access to the high performance computing facility Hawk-AI at HLRS, Stuttgart. We would like to express our gratitude for the valuable discussions and support provided by Prof. Nikolaus Adams, Deniz Bezgin, and Aaron Buhendwa from TU München.
%The study is Funded by Deutsche Forschungsgemeinschaft (DFG, German Research Foundation) under Germany's Excellence Strategy - EXC 2075 – 390740016. We acknowledge the support by the Stuttgart Center for Simulation Science (SimTech).
\end{acknowledgements}

\appendix

\section{Validation of DNS channel flow}\label{app:mean}

\begin{figure}
    \centering
    \includegraphics[width=1\linewidth]{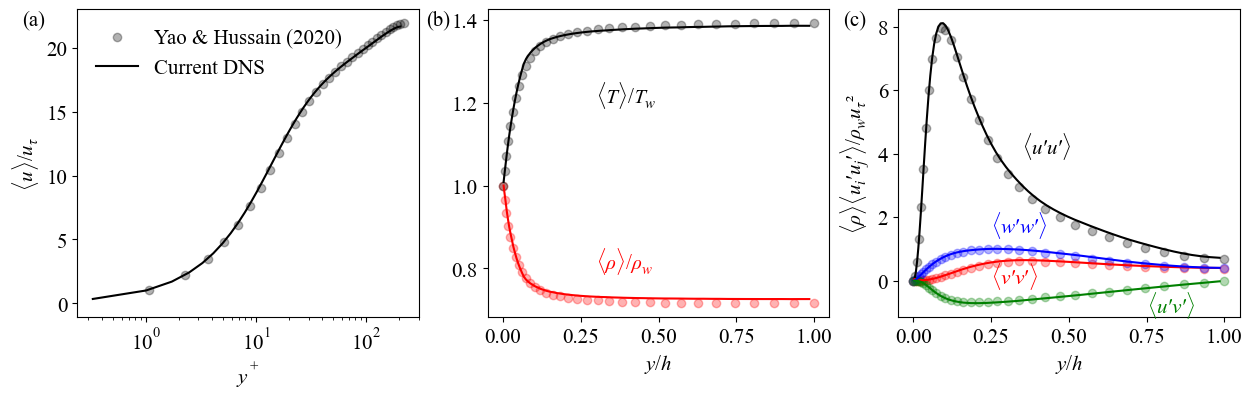}
    \caption{Comparison of mean statistics between the current DNS and \cite{yao2021drag}. (a) mean streamwise velocity $\langle u\rangle$;(b) mean temperature $\langle T\rangle$ and mean density $\langle \rho\rangle$; (c) turbulent intensity $\langle u_i'u_i'\rangle$ and Reynolds stress $\langle u'v'\rangle$.}
    \label{validation1}
\end{figure}

Figure \ref{validation1} shows a comparison of the current DNS results with those of \cite{yao2021drag} for validation of basic turbulent flow statistics. Yao's DNS had a resolution of 512×129×256, with a domain size of \(6\pi  \times 2 \times 2\pi  \), which is larger in the streamwise and spanwise directions, and features a higher resolution than the current DNS setup. Despite these differences, figure \ref{validation1}(a) shows that the mean streamwise velocity profile \( \langle u \rangle / u_{\tau} \), plotted against the wall-normal coordinate \( y^+ \), demonstrates excellent agreement between the current DNS (solid black line) and the reference data from \cite{yao2021drag} (gray circles).
In figure \ref{validation1}(b), the temperature \( \langle T \rangle / T_w \) and density \( \langle \rho \rangle / \rho_w \) profiles are shown as functions of the normalized wall-normal distance \( y/h \). The current DNS captures the near-wall temperature peak and the density decay towards the centerline accurately, aligning well with the reference data.
Figure \ref{validation1}(c) compares the Reynolds stress components $\langle u'_iu'_j \rangle$ between the current DNS (solid lines) and the reference DNS (circle markers). Strong agreement is observed across all components, demonstrating that the current DNS captures the key turbulent stress distributions despite the differences in resolution and domain size. This comparison confirms the reliability of the current DNS in reproducing the essential turbulent flow features and stress profiles.

\section{Validation of automatic differentiation with finite difference }

\begin{figure}
    \centering
    \includegraphics[width=1\linewidth]{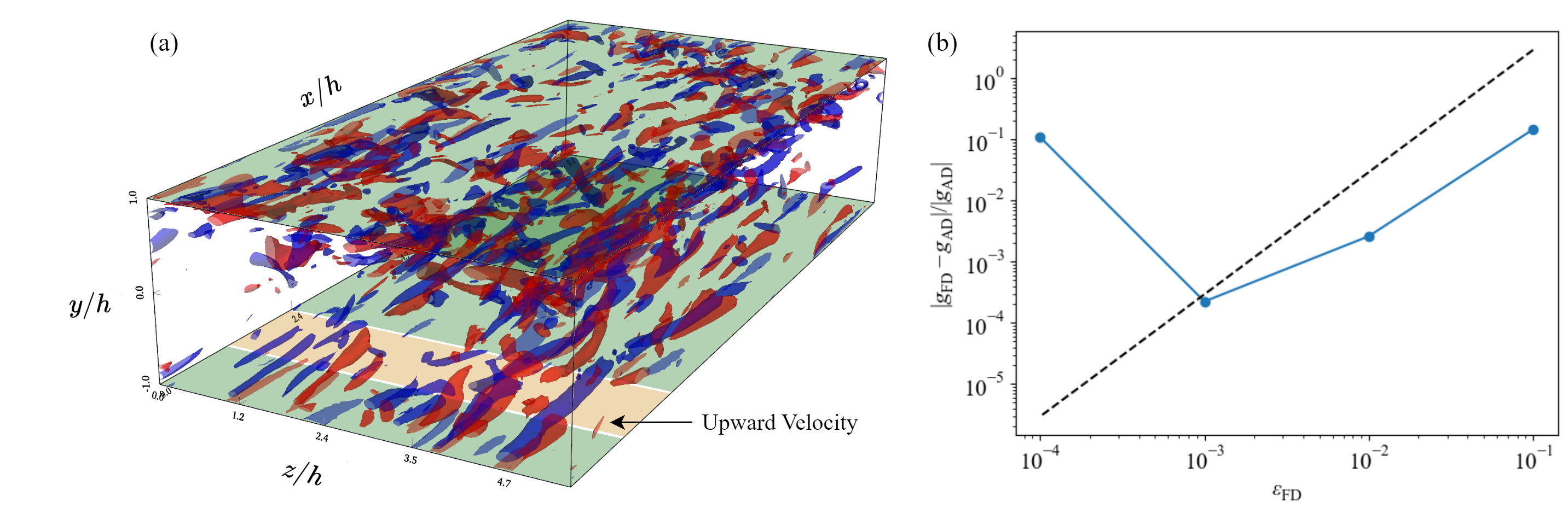}
    \caption{(a) Opposition control setup for AD validation. The green regions near the upper and lower walls represent areas where the control amplitude is set to zero, while the orange regions denote areas with a uniform control amplitude $\phi$. (b) Convergence of finite-difference gradients towards automatic differentiation gradients.. The blue line represents the relative error between FD and AD gradients as a function of the step size. The dashed line indicates second-order convergence.}
    \label{fig:AD-vali}
\end{figure}
We validate automatic differentiation (AD) using finite difference (FD) method in the context of opposition control. 
Figure \ref{fig:AD-vali}(a) shows the case setup. The green areas in the upper and lower walls represent regions where the opposition control amplitude is set to zero. The orange regions denote areas where the control amplitude is assigned a finite uniform value $\alpha$. We evaluated the sensitivity of the cumulative TKE over a horizon of $\Delta t^+ = 50$ with respect to the amplitude $\alpha$ when $\alpha=0$ using both AD and FD methods.

Note that instead of computing gradient at a point on the control surface,  we compute the gradient over an area of uniform control input. This lies in the nature of the methods being compared.  While AD can compute gradients accurately even with a control input applied at a single point, FD methods struggle with this due to round-off errors. When using a very small control input like a single point, the changes in the output can be too subtle, causing FD to become unreliable because it cannot distinguish these small variations amidst numerical noise.

Figure \ref{fig:AD-vali}(b) displays the comparison between the derivatives computed through AD and those obtained using the FD approach. Specifically, we use a second-order central difference scheme where the amplitude is incremented by \(\epsilon_{\mathrm{FD}}\), leading to a derivative estimate through 
\begin{equation}
     g_{\mathrm{FD}}=(k|_{\alpha = \varepsilon_{\mathrm{FD}}} - k|_{\alpha= -\varepsilon_{\mathrm{FD}}}) / 2\varepsilon_{\mathrm{FD}} .
\end{equation}
 The blue line represents the relative error between the gradient calculated using FD ($g_{\mathrm{FD}}$) and AD ($g_{\mathrm{AD}}$) as the step size $\varepsilon_{\mathrm{FD}}$ is reduced. It is evident from the plot that as $\varepsilon_{\mathrm{FD}}$ decreases, the error diminishes, showing the expected trend of improved accuracy with smaller step sizes. The dashed line indicates second-order convergence, which serves as a reference for the behavior of the FD error. The trend of the blue line aligns well with this reference initially, confirming the second-order accuracy of the FD method.

However, as the step size $\varepsilon_{\mathrm{FD}}$ decreases further, numerical errors, such as round-off errors, begin to dominate, causing the deviation from the dashed line. This underscores the advantage of AD, which, unlike FD, is not limited by the choice of a step size and provides exact gradients up to machine precision. 

\section{Decomposition of mean friction in permeable wall channels}\label{app:RD}
Considering the channel flow, we assume statistical homogeneity in the spanwise and streamwise directions and
 symmetry with respect to the central plane of the channel. The compressible Reynolds averaged momentum equation in the streamwise ($x$-)direction is
 \begin{equation}
     \frac{\partial\langle\rho u\rangle}{\partial t}+\frac{\partial\langle\rho u v\rangle}{\partial y}=\frac{\partial\left\langle\tau_{y x}\right\rangle}{\partial y}+\langle\rho f\rangle
     \label{eq:um}
 \end{equation}
 where $u$ and $v$ are respectively the streamwise and wall-normal components of the
 transient velocity, $\rho$ is density, $t$ is time and $\tau_{yx}$ is the shear stress in the streamwise
 direction. A uniform body force $f$ is added to drive the flow in the streamwise
 direction. Integrating (\ref{eq:um}) from the wall surface to the central
 plane gives 

 \begin{equation}
     \frac{\partial Q}{\partial t}=-\tau_w+\rho_b h f
     \label{eq:um-f}
 \end{equation}
 where $Q$ is the mass flow rate across the traverse plane. In current study, $Q$ is controlled to be constant, hence $f=\tau_w/\rho_bh$. 

Rewriting the left-hand side of (\ref{eq:um}) in the form of Favre average gives 
\begin{equation}
    \frac{\partial\langle\rho u\rangle}{\partial t}+\frac{\partial\langle\rho u v\rangle}{\partial y}=\langle\rho\rangle\frac{\partial\{u\}}{\partial t}+\frac{\partial\langle\rho\rangle\{u\}\{v\}}{\partial y}+\frac{\partial\langle\rho\rangle\left\{u^{\prime \prime} v^{\prime \prime}\right\}}{\partial y}
    \label{eq:um-r}
\end{equation}
For no-slip boundary condition and zero-net-flux opposition control boundary condition, the second term on the right hand side, representing the mean convection in vertical direction, is considered as zero since $\{v\}=0$. However, for the permeable wall condition, this term may not be zero as flux through the wall is allowed. 

Substituting (\ref{eq:um-f}) and (\ref{eq:um-r}) into (\ref{eq:um}), we have

\begin{equation}
    \langle\rho\rangle \frac{\partial\{u\}}{\partial t}=-\frac{\partial\langle\rho\rangle\{u\}\{v\}}{\partial y}-\frac{\partial\langle\rho\rangle\left\{u^{\prime \prime} v^{\prime \prime}\right\}}{\partial y}+\frac{\partial\left\langle\tau_{y x}\right\rangle}{\partial y}+\frac{\langle\rho\rangle}{\rho_b h} \tau_w .
    \label{eq:u_tilde}
\end{equation}

We convert the original reference frame, which is stationary with respect to the wall, into an absolute reference frame, traveling at velocity $u_b$. Consequently, the friction on the wall generates a non-zero power that contributes to the mean kinetic-energy budget. Let the subscript $a$ represent the variables in the absolute frame. We have

\begin{equation}
    t_a=t, \quad \rho_a=\rho, \quad x_a=x-u_b t, \quad y_a=y, \quad u_a=u-u_b, \quad v_a=v
    \label{eq:frame}
\end{equation}

Substituting (\ref{eq:frame}) into (\ref{eq:u_tilde}) yields

\begin{equation}
    \langle\rho_a\rangle \frac{\partial\{u_a\}}{\partial t_a}=-\frac{\partial\langle\rho_a\rangle\{u_a\}\{v_a\}}{\partial y_a}-\frac{\partial\langle\rho_a\rangle\left\{u_a^{\prime \prime} v_a^{\prime \prime}\right\}}{\partial y_a}+\frac{\partial\left\langle\tau_{y x}\right\rangle}{\partial y_a}+\frac{\langle\rho_a\rangle}{\rho_b h} \tau_w .
    \label{eq:u_tilde2}
\end{equation}

Multiplying both sides of (\ref{eq:u_tilde2}) by $\{u_a\}$, we have the energy budget equation 

 \begin{equation}
     \left\langle\rho_a\right\rangle \frac{\partial\left\{K_a\right\}}{\partial t_a}=-\left\{u_a\right\}\frac{\partial\langle\rho_a\rangle\{u_a\}\{v_a\}}{\partial y}-\left\{u_a\right\} \frac{\partial\left\langle\rho_a\right\rangle\left\{u_a^{\prime \prime} v_a^{\prime \prime}\right\}}{\partial y_a}+\left\{u_a\right\} \frac{\partial\left\langle\tau_{y x}\right\rangle}{\partial y_a}+\left\{u_a\right\} \frac{\left\langle\rho_a\right\rangle}{\rho_b h} \tau_w .
     \label{eq:budget}
 \end{equation}
Where $K_a=\{u_a\}^2/2$ is the averaged streamwise kinetic energy of unit mass in the absolute frame. For channel flow statistically homogeneous in time, ${\partial K_a}/{\partial t_a}=0$. A single integration over the half-channel is then performed on (\ref{eq:budget}), using symmetry boundary conditions at the centre line and $\{u_a\}|_{y=0}=-u_b$
 leads to the formula of the skin-friction coefficient in the absolute reference frame,

\begin{equation}
\begin{split}
        C_f&=\frac{2 \tau_w}{\rho_b u_b^2}\\
        &=\underbrace{\frac{2}{\rho_b u_b^3} \int_0^h\left\langle\tau_{y x}\right\rangle \frac{\partial\left\{u_a\right\}}{\partial y_a} \mathrm{~d} y_a}_{C_{f}^\nu}+\underbrace{\frac{2}{\rho_b u_b^3} \int_0^h\left\langle\rho_a\right\rangle\left\{-u_a^{\prime \prime} v_a^{\prime \prime}\right\} \frac{\partial\left\{u_a\right\}}{\partial y_a} \mathrm{~d} y_a}_{C_{f}^T}\\
        &+\underbrace{\frac{2}{\rho_b u_b^3} \int_0^h\left\langle\rho_a\right\rangle\left(-\{u_a\}\{ v_a\}\right) \frac{\partial\{u_a\}}{\partial y_a} \mathrm{~d} y_a}_{C_{f1}^V}\underbrace{-\frac{2\rho_w \{ v_a\}_w}{\rho_b u_b}}_{C_{f2}^V}.
\end{split}
\label{eq:cfa}
\end{equation}

Rewriting (\ref{eq:cfa}) as a function of the usual (wall) reference frame variables yields

\begin{equation}
\begin{split}
       C_f=\underbrace{\frac{2}{\rho_b u_b^3} \int_0^h\left\langle\mu\left(\frac{\partial u}{\partial y}+\frac{\partial v}{\partial x}\right)\right\rangle \frac{\partial\{u\}}{\partial y} \mathrm{~d} y}_{C_{f}^\nu}+\underbrace{\frac{2}{\rho_b u_b^3} \int_0^h\langle\rho\rangle\left\{-u^{\prime \prime} v^{\prime \prime}\right\} \frac{\partial\{u\}}{\partial y} \mathrm{~d} y}_{C_{f }^T}\\
       \underbrace{-\frac{2}{\rho_b u_b^3} \int_0^h\langle\rho\rangle(\{u\}-u_b)\{v\}\frac{\partial\{u\}  }{\partial y} \mathrm{~d} y}_{C_{f1}^V}\underbrace{-\frac{2\rho_w \{ v\}_w}{\rho_b u_b}}_{C_{f2}^V}, 
\end{split}
\end{equation}
where 

\begin{equation}
    C_{f}^{\nu}=\underbrace{\frac{2}{\rho_b u_b^3} \int_0^h\langle\mu\rangle \frac{\partial\langle u\rangle}{\partial y} \frac{\partial\{u\}}{\partial y} \mathrm{~d} y}_{C_{f}^M}+\underbrace{\frac{2}{\rho_b u_b^3} \int_0^h\left\langle\mu^{\prime} \frac{\partial u^{\prime}}{\partial y}+\mu^{\prime} \frac{\partial v^{\prime}}{\partial x}\right\rangle \frac{\partial\{u\}}{\partial y} \mathrm{~d} y}_{C_{f}^F} .
\end{equation}
$C_{f}^{\nu}$ is associated with molecular viscosity dissipation, $C_f^M$ is the mean shear term; $C_f^F$ is the viscosity variation term; $C_f^T$ the turbulent-convection term; $C_{f1}^V$ and $C_{f1}^V$ mean convection terms that are generated due to the non-zero $\{v\}$ in the channel and at the wall. 

\bibliographystyle{jfm}
\bibliography{references.bib}

\end{document}